\documentclass[a4paper,11pt]{article}
\pdfoutput=1
\usepackage{dcolumn}
\usepackage{bm}
\usepackage{graphicx}
\usepackage{amssymb,amsmath}
\usepackage{multirow}
\usepackage{hhline}
\usepackage{cite,color,xcolor,url}
\def\linkcolor{cyan!70!black}

\usepackage[
colorlinks=true
,urlcolor=\linkcolor
,anchorcolor=\linkcolor
,citecolor=\linkcolor
,filecolor=\linkcolor
,linkcolor=\linkcolor
,menucolor=\linkcolor
,linktocpage=true
,pdfproducer=medialab
,pdfa=true
]{hyperref}

\usepackage{slashed}
\usepackage{epsfig,psfrag,rotating,soul}
\usepackage{rotfloat}
\usepackage[font={small}]{caption}
\usepackage{colortbl}
\usepackage{upgreek}

\evensidemargin \oddsidemargin
\newcommand{\be}{\begin{equation}}
\newcommand{\ee}{\end{equation}}

\newcommand{\MET}{E_T^\text{miss}}
\newcommand{\phimet}{\phi^\text{miss}}

\newcommand{\beq}{\begin{equation}} 
\newcommand{\eeq}{\end{equation}} 
\newcommand{\ba}{\begin{array}}  
\newcommand{\ea}{\end{array}} 
\newcommand{\bea}{\begin{eqnarray}}  
\newcommand{\eea}{\end{eqnarray} }  
\newcommand{\bal}{\begin{align}}
\newcommand{\eal}{\end{align}}   
\newcommand{\bi}{\begin{itemize}}  
\newcommand{\ei}{\end{itemize}}  
\newcommand{\ben}{\begin{enumerate}}  
\newcommand{\een}{\end{enumerate}}  
\newcommand{\bc}{\begin{center}}
\newcommand{\ec}{\end{center}} 
\newcommand{\bt}{\begin{table}}
\newcommand{\et}{\end{table}}  
\newcommand{\btb}{\begin{tabular}}
\newcommand{\etb}{\end{tabular}}

\newcommand{\fref}[1]{Figure~\ref{#1}} 
\newcommand{\eref}[1]{Eq.~(\ref{#1})} 
\newcommand{\tabref}[1]{Table~\ref{#1}}

\newcommand{\aref}[1]{Appendix~\ref{#1}}
\newcommand{\sref}[1]{Section~\ref{#1}}

\newcommand{\ttbbsemilep}{t\bar{t}\vert_{\rm semilep}b\bar{b}}
\newcommand{\ttbbsemitau}{t\bar{t}\vert_{\rm semitau}b\bar{b}}
\newcommand{\zbbbb}{Z\vert_{\rm inv}b\bar{b}b\bar{b}}

\allowdisplaybreaks

\let\OLDthebibliography\thebibliography
\renewcommand\thebibliography[1]{
  \OLDthebibliography{#1}
  \setlength{\parskip}{0pt}
  \setlength{\itemsep}{0pt plus 0.3ex}
}

\begin{document}

\begin{titlepage}

\thispagestyle{empty}

\def\thefootnote{\fnsymbol{footnote}}

\begin{flushright}
IFT-UAM/CSIC-23-60
\end{flushright}

\vspace*{1cm}

\begin{center}

\begin{Large}
\textbf{\textsc{
Machine-Learning Performance on \\[.25em] Higgs-Pair Production Associated with Dark Matter at the LHC
}}
\end{Large}

\vspace{1cm}

{\sc 
Ernesto~Arganda$^{1}$%
\footnote{{\tt \href{mailto:ernesto.arganda@uam.es}{ernesto.arganda@uam.es}}}%
, Manuel~Epele$^{2}$%
\footnote{{\tt \href{mailto:manuepele@fisica.unlp.edu.ar}{manuepele@fisica.unlp.edu.ar}}}%
, Nicolas~I.~Mileo$^{2}$%
\footnote{{\tt \href{mailto:mileo@fisica.unlp.edu.ar}{mileo@fisica.unlp.edu.ar}}}%
, and Roberto~A.~Morales$^{2}$%
\footnote{{\tt \href{mailto:roberto.morales@fisica.unlp.edu.ar}{roberto.morales@fisica.unlp.edu.ar}}}%
}

\vspace{0.5truecm}

{\sl

$^1$Departamento de Física Teórica and Instituto de F\'{\i}sica Te\'orica UAM-CSIC, \\
Universidad Autónoma de Madrid, Cantoblanco, 28049 Madrid, Spain

\vspace*{0.1cm}

$^2$IFLP, CONICET - Dpto. de F\'{\i}sica, Universidad Nacional de La Plata, \\ 
C.C. 67, 1900 La Plata, Argentina

\vspace*{0.1cm}
}

\vspace*{2mm}

\end{center}

\vspace{0.1cm}

\renewcommand*{\thefootnote}{\arabic{footnote}}
\setcounter{footnote}{0}

\begin{abstract}
\noindent
Di-Higgs production at the LHC  associated with missing transverse energy is explored in the context of simplified models that generically parameterize a large class of models with heavy scalars and dark matter candidates.
Our aim is to figure out the improvement capability of machine-learning tools over traditional cut-based analyses. In particular, boosted decision trees and neural networks are implemented in order to determine the parameter space that can be tested at the LHC demanding four $b$-jets and large missing energy in the final state. We present a performance comparison between both machine-learning algorithms, based on the maximum significance reached, by feeding them with different sets of kinematic features corresponding to the LHC at a center-of-mass energy of 14 TeV. Both algorithms present very similar performances and substantially improve traditional analyses, being sensitive to most of the parameter space considered for a total integrated luminosity of 1 ab$^{-1}$, with significances at the evidence level, and even at the discovery level, depending on the masses of the new heavy scalars. A more conservative approach with systematic uncertainties on the background of 30\% has also been contemplated, again providing very promising significances.

\end{abstract}


\end{titlepage}

\tableofcontents

\section{Introduction}
\label{intro}

In the last few years machine learning (ML) has become an standard and basic tool for experimental and phenomenological high-energy physics studies (for reviews see, for instance,~\cite{Larkoski:2017jix,Guest:2018yhq,Albertsson:2018maf,Radovic:2018dip,Carleo:2019ptp,Bourilkov:2019yoi,Feickert:2021ajf,Schwartz:2021ftp,Karagiorgi:2021ngt,Shanahan:2022ifi,Plehn:2022ftl}). Indeed, ML may be crucial to take full advantage of the data collected at the LHC in order to probe the standard model (SM) and new physics. In this sense, a key issue is whether ML techniques could replace traditional counting methods based on rectangular cuts. The ATLAS and CMS Collaborations have shown in many experimental analyses the potential of ML tools, as boosted decision trees (BDT) and neural networks (NN), to improve the LHC sensitivity, understood as the signal-to-background ratio, to beyond the SM (BSM) physics~\cite{CMS:2013hhn,ATLAS:2014bba,ATLAS:2015eiz,CMS:2017rpp,CMS:2017wtu,CMS:2018hnq,ATLAS:2018mpo,ATLAS:2019ebv,ATLAS:2019qrr,CMS:2019eln,CMS:2019rlz,ATLAS:2019tkk,CMS:2019dqq,CMS:2020cga,CMS:2020poo,ATLAS:2020lks,ATLAS:2020azv,CMS:2020tkr,CMS:2021beq,ATLAS:2021fbt,ATLAS:2022zhj,CMS:2022wjj,ATLAS:2022izj,ATLAS:2022zuc,CMS:2022dbt,CMS:2022hjj,CMS:2022pjv,CMS:2022idi,CMS:2022wjc,ATLAS:2023qzf,CMS:2023ktc,CMS:2023agg,CMS:2023pte,ATLAS:2023azi,ATLAS:2023ixc,CMS:2023bdh,CMS:2023mny,CMS:2023hwl,CMS:2023boe,CMS:2023dof,ATLAS:2023gzn,CMS:2023eos,CMS:2023jgd,CMS:2023xpx}. The main goal of this work is to find out the improvement capability of modern ML tools over cut-based analyses applied to a case study of physical interest: the production at the LHC of Higgs boson pairs associated with dark matter (DM) particles.

After the Higgs boson discovery~\cite{ATLAS:2012yve,CMS:2012qbp}, with a mass value of $m_h$ = 125.09 $\pm$ 0.24 GeV~\cite{ATLAS:2015yey} and which seems to be the scalar Higgs boson predicted by the SM~\cite{ATLAS:2016neq}, the search for extended Higgs sectors represents an intensive experimental program carried out at the LHC by ATLAS and CMS (for recent analyses see, for instance,~\cite{ATLAS:2017uhp,ATLAS:2018gfm,ATLAS:2019tpq,CMS:2019pzc,ATLAS:2020zms,CMS:2020efq,ATLAS:2020tlo,ATLAS:2021upq,CMS:2021far,CMS:2021aly,CMS:2022bcb,CMS:2022uox,CMS:2022kdx,CMS:2022tgk,ATLAS:2022enb,CMS:2022jqc,CMS:2022goy,ATLAS:2022fpx,ATLAS:2022rws,ATLAS:2022eap,CMS:2023pte,ATLAS:2023tkl,ATLAS:2023tlp,ATLAS:2023wqy,CMS:2023boe,ATLAS:2023zkt,ATLAS:2023vdy,ATLAS:2024itc}). Interestingly, these additional Higgs bosons may serve as portals to dark sectors~\cite{Silveira:1985rk,Schabinger:2005ei,Patt:2006fw,OConnell:2006rsp,LopezHonorez:2006gr,March-Russell:2008lng,Englert:2011yb,No:2015xqa,Brivio:2015kia,Ghorbani:2017qwf,Tunney:2017yfp,Bell:2017rgi,Yang:2017zor,Ghosh:2020fdc,Aguilar-Saavedra:2022xrb,Arcadi:2022lpp,PerezAdan:2023phe,Wang:2023suf,Carrasco:2023loy}, which could manifest in multi-Higgs final states with a large amount of missing transverse energy ($E_T^\text{miss}$), coming from the potential emission of DM that escapes from the LHC detectors. In fact, ATLAS and CMS are also conducting many searches for di-Higgs production plus $E_T^\text{miss}$~\cite{CMS:2017nin,CMS:2017may,ATLAS:2018tti,CMS:2019pov,ATLAS:2021yqv,CMS:2022vpy,CMS:2022spe,CMS:2022kdx,CMS:2023ehl,ATLAS:2023reg,ATLAS:2023act}.

In the context of these di-Higgs + $E_T^\text{miss}$ searches at the LHC, numerous phenomenological collider analyses have emerged~\cite{Matchev:1999ft,Kang:2015nga,Etesami:2015caa,Kang:2015uoc,Biswas:2016ffy,Banerjee:2016nzb,Brivio:2017ije,Arganda:2017wjh,Chen:2018dyq,Bernreuther:2018nat,Titterton:2018pba,Blanke:2019hpe,Alves:2019emf,Flores:2019hcf}. Based on the search strategies developed in~\cite{Arganda:2017wjh} and motivated by the use of multivariate analysis (MVA) tools in~\cite{Blanke:2019hpe} and~\cite{Flores:2019hcf}~\footnote{During the completion of this work, Ref.~\cite{Hammad:2023sbd} appeared in which similar aspects are discussed with multi-scale cross attention transformer encoders for the di-Higgs boson production, but without the associated DM production. That is, only the final state composed of four $b$-jets is considered, without a large amount of $\MET$.}, in this work we aim to study the performance of modern ML algorithms on this type of phenomenological analyses. Concretely, in order to try to increase the LHC sensitivity to di-Higgs + $E_T^\text{miss}$ signatures in general frameworks of extended Higgs sectors, we will employ the {\tt XGBoost} toolkit~\cite{Chen:2016btl,Chen:2016:XST:2939672.2939785} and deep neural networks (DNN)~\cite{bishop1995neural,10.1162/neco.2006.18.7.1527,doi:10.1126/science.1127647,SCHMIDHUBER201585}.

In particular, we work within a general framework defined by simplified models based on effective couplings and interaction terms, inspired by~\cite{Blanke:2019hpe}, and consider the resonant production of a pair of heavy scalars at the LHC, with one of the heavy scalar decaying invisibly to a pair of DM particles, and the other one decays to two SM-like Higgs bosons, which subsequently decay to bottom-quark pairs as in~\cite{Arganda:2017wjh}. Therefore, the LHC signature under study consists of 4 $b$-jets plus a large amount of $E_T^\text{miss}$, whose dominant irreducible backgrounds are $Z b \bar b b \bar b$ and $t \bar t b \bar b$.
We will define several benchmarks, covering the region of the parameter space of interest, with different kinematic characteristics, and train the ML algorithms with balanced signal and background event samples for each. 
We will maximize the signal significance for each benchmark, including a rough estimation of the systematic uncertainties, scanning over the different values of the ML classifier scores. We will also compare the performances of {\tt XGBoost} and DNN with respect to traditional cut-based search strategies.

The paper is structured as follows: in Section~\ref{ph-frame} we introduce the simplified BSM model that produces the LHC signature under study, define the signal benchmarks and present the collider variables with which we feed the ML algorithms to discriminate between signal and background; Section~\ref{MLalgo} is devoted to summarize the main features of {\tt XGBoost} and DNN algorithms used for differentiating signal to background; Section~\ref{res} is dedicated to the analysis of the {\tt XGBoost} and DNN performances compared to the rectangular cut-based collider analyses; finally, we leave Section~\ref{conclu} to discuss our main conclusions.

\section{Phenomenological Framework}
\label{ph-frame}

\begin{figure}[t!]
\begin{center}
\includegraphics[width=0.65\textwidth]{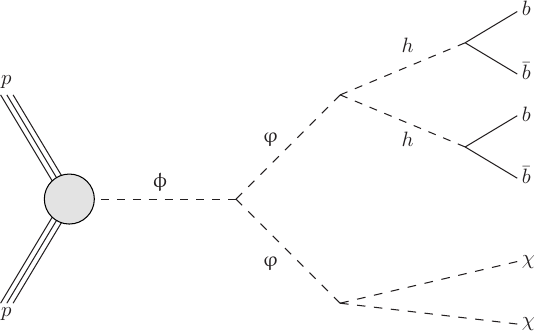}
\caption{Decay chain topology of LHC production of a heavy scalar $\upphi$, that decays into a pair of intermediate scalars $\upvarphi$. One of the scalar $\upvarphi$ decays into a pair of SM-like Higgs bosons $h$, which subsequently decay into $b$-quark pairs, whilst the other scalar $\upvarphi$ decays directly into a pair of DM candidates, the lightest scalar $\chi$.}\label{ResonantTopology}
\end{center}
\end{figure}

We start our analysis of di-Higgs + $\MET$ production at the LHC, with a center-of-mass energy of 14 TeV and a luminosity of $\cal{L}$ = 1 ab$^{-1}$, in the context of a general class of simplified models, inspired by the effective couplings and interactions terms presented in~\cite{Blanke:2019hpe}. 
This class of models relies on an extended scalar sector with three real scalar particles. The heaviest of these new scalars, $\upphi$, is produced via gluon fusion at the LHC (through an effective dimension-five interaction) and predominantly decays to a pair of intermediate scalars $\upvarphi$. 
This scalar interacts with the visible sector only through its coupling with the SM Higgs boson $h$.
The lightest scalar $\chi$ is the DM candidate within the framework of this effective field theory. 
Although a detailed study of the correct relic density and the bounds from DM direct and indirect detection experiments is out of the scope of this work, we would like to notice that these limits would not constrain the LHC phenomenology presented in this work~\cite{Blanke:2019hpe}.
We consider only the resonant topology $pp\to \upphi\to \upvarphi\upvarphi\to (hh)(\chi\chi)$ since it is the most pessimistic scenario presented in~\cite{Blanke:2019hpe} and the necessity of ML algorithms is well motivated. This topology is shown in \fref{ResonantTopology} and is also very similar to one studied in our previous work~\cite{Arganda:2017wjh}. 
In particular we consider a variety of benchmarks of this class of simplified models scanning the mass $m_\upphi$ of the heavy scalar in the range [750, 1500] GeV and the mass $m_\upvarphi$ of the intermediate state in the range [275, $m_\upphi/2$] GeV~\footnote{We choose these values for the masses of the new scalars because of their potential accessibility to the LHC energy scales and their phenomenological interest. There are no specific ATLAS and CMS searches for our proposed new physics signal~\cite{ATLAS:2023xhq,EPS-HEP2023DMsearchesCMS}, so the exclusion limits on this parameter space are very poor. However, if limits existed that were in tension with some of the selected benchmarks, which is not the case, it would also not be of concern since the benchmarks have been chosen as a case study with the main goal of weighting the ML performance for this type of LHC final states.}. Also, we set $m_{\chi}=25$ GeV for all the benchmarks since it has no impact on the phenomenology of the considered topology.

The interaction Lagrangian for this simplified model is given by
\begin{equation}
    \mathcal{L} = \frac{C_{\upphi gg}}{\Lambda}\upphi G_{\mu\nu}G^{\mu\nu} +\frac{m_{\upphi\upvarphi\upvarphi}}{2}\upphi\upvarphi\upvarphi + \frac{m_{\upvarphi hh}}{2}\upvarphi hh +\frac{m_{\upvarphi\chi\chi}}{2}\upvarphi\chi\chi \,,
\end{equation}
and the resulting cross section for the topology in \fref{ResonantTopology} is factorized as
\begin{equation}
    \sigma(pp\to b\bar{b}b\bar{b}\chi\chi) = \sigma(pp\to\upphi) \cdot \text{BR}(\upphi\to\upvarphi\upvarphi) \cdot 2\cdot \text{BR}(\upvarphi\to hh) \cdot \text{BR}(\upvarphi\to\chi\chi) \cdot \left[\text{BR}(h\to b\bar{b})\right]^2 \,.
    \label{resonantxs}
\end{equation}
Following~\cite{Blanke:2019hpe}, the effective coupling $C_{\upphi gg}$ collects the effect of heavy quarks at scale $\Lambda=1$ TeV (well above the EW scale $v=246$ GeV) in an UV complete theory and the  production cross section $\sigma(pp\to\upphi)$ can be obtained from the LHC Higgs cross-section working group for BSM scalars $S$~\cite{CERNYellowReportPageBSMAt14TeV}:
\begin{equation}
    \sigma(pp\to\upphi) = \left(\frac{v}{1\,\text{TeV}}\right)^2\sigma(pp\to S) \,.
\end{equation}
We are interested in the kinematic region corresponding to the mass of the heavy scalar larger than twice the mass of the intermediate scalar. Then the two decay channels for the heavy scalar are
\begin{equation}
    \Gamma(\upphi\to\upvarphi\upvarphi)=\frac{m_{\upphi\upvarphi\upvarphi}^2}{32\pi m_\upphi}\sqrt{1-\frac{4m_\upvarphi^2}{m_\upphi^2}} \quad\text{and}\quad \Gamma(\upphi\to gg)=\frac{C_{\upphi gg}^2}{1\,\text{TeV}^2}\frac{2m_\upphi^3}{\pi} \,.
\end{equation}
Setting the coupling $m_{\upphi\upvarphi\upvarphi}$ equal to the EW scale $v=246$ GeV and far to the $\upvarphi\upvarphi$ threshold, we have BR$(\upphi\to\upvarphi\upvarphi)\sim 1$.

In addition, the couplings $m_{\upvarphi hh}$ and $m_{\upvarphi\chi\chi}$ are such that the branching ratios of $\upvarphi$ decaying to $hh$ and $\chi\chi$ maximize their product. Concretely, we fix BR$(\upvarphi\to hh)$ = BR$(\upvarphi\to\chi\chi)=0.5$. We consider also BR$(h\to b\bar{b})$ = 0.58~\cite{LHCHiggsCrossSectionWorkingGroup:2016ypw}.

Signal and background events are generated with {\tt MadGraph\_aMC@NLO 2.8.1}~\cite{Alwall:2014hca}, the parton showering and hadronization is carried out by using {\tt PYTHIA 8.2}~\cite{Sjostrand:2014zea}, and the detector response is simulated through {\tt Delphes 3.3.3}~\cite{deFavereau:2013fsa,Selvaggi:2014mya,Mertens:2015kba}, with the jets being reconstructed with {\tt FastJet 3.4.0}~\cite{Cacciari:2005hq,Cacciari:2011ma}. We use the default setup for the ATLAS detector provided by {\tt Delphes 3.3.3}, in which the $b$-tagging efficiency, and the light and charm jet mistag rates are given by
\begin{equation}
    \label{eff}
    \epsilon_b = \frac{24\tanh (0.003\, p_T)}{1+0.086\, p_T} \,,\quad \epsilon_j=0.002+ 7.3\times 10^{-6}\,p_T \,, \quad \epsilon_c = \frac{0.2\,\tanh (0.02\, p_T)}{1+0.0034\,p_T} \,,
\end{equation}
respectively. With this working point a maximum $b$-tagging efficiency of $\sim 73\%$ is reached for a jet transverse momentum of $p_T\sim 120$ GeV.
We work in the 4-flavor scheme and we do not apply jet matching since we do not have light jets in the hard process of our signal.
The simulation input files and the internal analysis codes are available upon request to the authors.

We focus on a signal region defined by imposing the following cuts at detector level:
\begin{equation}
\label{SRdetcuts}
N_b=4\,,\quad N_{lep}=0 \,, \quad p_T^{j,b} > 20 \, \text{GeV} \,, \quad \MET > 200 \, \text{GeV} \,,
\end{equation}
where $N_b$ and $N_{lep}$ are the number of $b$-jets and leptons (electrons and muons), respectively, $p_T^{j,b}$ is the transverse momentum of light and $b$-tagged jets, and $\MET$ is the missing transverse energy.  
In particular, the lepton-veto disfavors the presence of missing energy coming from neutrinos in the top-quark decays.

\begin{figure}[t!]
\begin{center}
\includegraphics[width=0.7\textwidth]{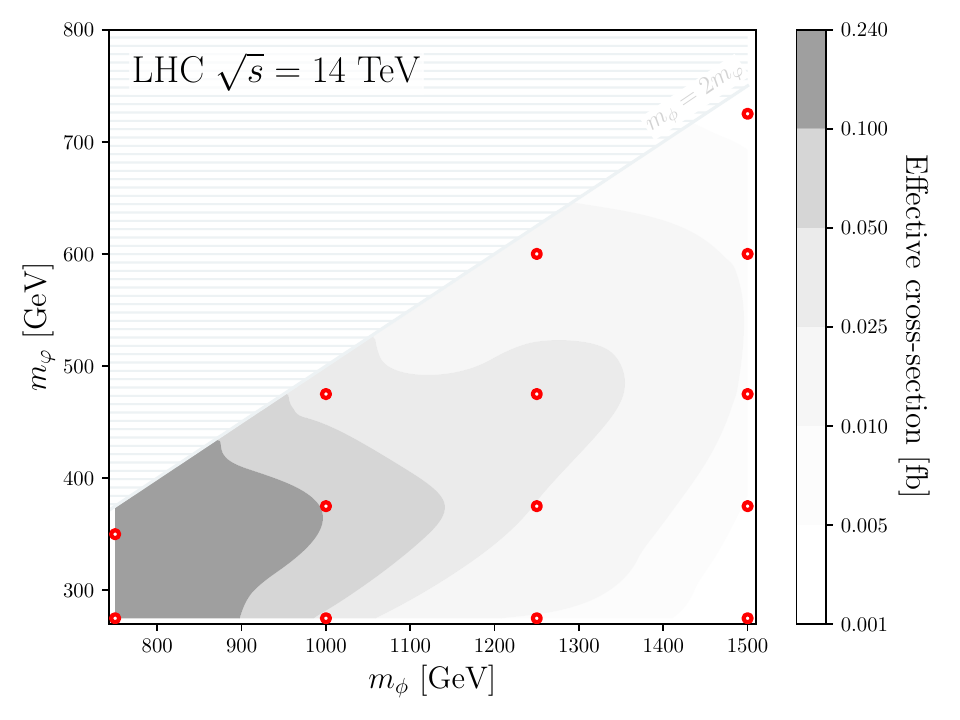}
\caption{Effective cross section in the plane $[m_\upphi,m_\upvarphi]$ defined as $\sigma_\text{eff}=\sigma(pp\to b\bar{b}b\bar{b}\chi\chi)\cdot\varepsilon_\text{SR}$. The region with $m_\upphi<2m_\upvarphi$ is kinematically closed. Red points correspond to the benchmarks with events simulated for the training (see Table~\ref{tab:effxs-sig} of Appendix~\ref{App-xs} for the specific values of the effective cross-section of each benchmark).}
\label{fig-effxs}
\end{center}
\end{figure}

For each benchmark, the acceptance $\varepsilon_\text{SR}$ is the fraction of the simulated signal events which satisfy \eref{SRdetcuts}. 
Combining this quantity with \eref{resonantxs}, the effective cross section is defined as $\sigma_\text{eff} = \sigma(pp\to b\bar{b}b\bar{b}\chi\chi)\cdot\varepsilon_\text{SR}$ and, after multiplying by the luminosity, it represents the number of expected signal events in the signal region of \eref{SRdetcuts}. 
The effective cross sections for the 14 simulated benchmarks are shown in \fref{fig-effxs} by red points in the plane $[m_\upphi,m_\upvarphi]$.
For fixed heavy scalar mass $m_\upphi$, the cross section is roughly $m_\upvarphi$ independent. However, the acceptance $\varepsilon_\text{SR}$ becomes smaller for low $m_{\upvarphi}$ and also close to the threshold, $m_\upvarphi\sim m_\upphi/2$, leading to the profile of the effective cross section observed in this figure. In particular, near the threshold, both scalars $\upvarphi$ are produced almost at rest, which strongly reduces the acceptance $\epsilon_{\mathrm{SR}}$ due to the impact of the missing transverse energy cut. On the other hand, when $m_{\upvarphi}$ is close to the production threshold of two Higgs bosons from the decay of $\upvarphi$, these emerge almost at rest in the $\upvarphi$-frame and then they are emitted roughly along the direction of the momentum of $\upvarphi$ in the collision frame, which make their decay products more difficult to resolve. Thus, the acceptance $\epsilon_{\mathrm{SR}}$ decreases significantly when demanding four $b$-jets. As an example, the acceptance of requiring $N_b=4$ drops from 0.9 to 0.02 when going from benchmark 1250\_475\_25 to 1250\_275\_25~\footnote{From now on the benchmark $m_\upphi$\_$m_\upvarphi$\_$m_\chi$ corresponds to masses of the heavy, intermediate and dark matter scalars equal to $m_\upphi$, $m_\upvarphi$ and $m_\chi$, respectively.}.

Given the topology of the final state and the cuts introduced in Eq.~(\ref{SRdetcuts}), the dominant irreducible SM backgrounds are $Z+b\bar{b}b\bar{b}$, with the invisible decay of the $Z$ boson ($Z\to\nu\bar{\nu}$), and $t\bar{t}+b\bar{b}$, with the semi-leptonic decay modes $t(\to b jj)\bar{t}(\to \bar{b}\ell^- \bar{\nu})$ and $t(\to b \ell^+ \nu)\bar{t}(\to \bar{b} j j)$ of the top-antitop pair. From now on, we denote them as $\zbbbb$, and $\ttbbsemilep$ and  $\ttbbsemitau$ for the semi-leptonic modes of the $t\bar{t}b\bar{b}$ background with $\ell=e,\mu$ and $\ell=\tau$, respectively. 
On the other hand, the major reducible backgrounds correspond to $t\bar{t}$ and $V$+jets (concretely $Z+b\bar{b}+jj$, $Z+jjjj$, $W^\pm+b\bar{b}+jj$ and $W^\pm+jjjj$).
Notice that the main source of missing transverse energy for these backgrounds are the neutrinos coming from the gauge bosons and $t\bar{t}$ decay channels. In this regard, we safely neglect the QCD multijet production since it does not have a genuine source of missing transverse energy.

Therefore, with the intention of reducing the large background cross sections and making event generation more efficient, we impose the following generator-level cuts on the $p_T$ of the light jets and $b$-jets and also to the missing transverse energy~\footnote{At generator-level, it corresponds to the sum of neutrino's momenta which are present in these background processes. At detector-level, this is the main source of $\MET$ for backgrounds. Notice that in the signal case, it comes from the invisible decay $\upvarphi\to\chi\chi$.} for the background simulation:
\begin{equation}
p_T^{b} > 20 \, \text{GeV} \,, \quad p_T^{j} > 20 \, \text{GeV} \,, \quad \MET > 200 \, \text{GeV}  \,.
\label{generatorcuts}
\end{equation}

Taking this into account we only keep the three dominant irreducible backgrounds and two of the reducible $V$+jets category. 
With the cuts in \eref{generatorcuts}, the relevant cross sections are
\bea
&&\sigma(\ttbbsemilep) = 33.4 \,\text{fb} \,, \quad \sigma(\ttbbsemitau) = 16.7 \,\text{fb} \,, \quad \sigma(\zbbbb) = 3.7 \,\text{fb}  \,,  \nonumber\\
&&\sigma(Z\vert_{\rm inv}b\bar{b}jj) = 937 \,\text{fb} \,, \quad \sigma(W^\pm\vert_{\rm semilep}b\bar{b}jj) = 373 \,\text{fb} \,.
\label{backgroundxs}
\eea

With all the above considerations, we focus our ML analysis in a particular signal region demanding at detector level the \eref{SRdetcuts}.
The effective cross-sections for the relevant backgrounds in this signal region can be found in Table~\ref{tab:effxs-bkg} of Appendix~\ref{App-xs}.
The detector-level events, for signal and background, are then converted into data arrays, which are to be used as inputs for the ML algorithms, as follows:
\begin{itemize}

\item We reconstruct the Higgs bosons from the two $b \bar b$ pairs that minimize
\begin{equation}
\chi_{hh}=\sqrt{\left(\frac{m_{2b}^{\rm lead}-m_h}{0.1m_{2b}^{\rm lead}}\right)^2+\left(\frac{m_{2b}^{\rm subl}-m_h}{0.1m_{2b}^{\rm subl}}\right)^2} \,,
\label{chiHH}
\end{equation}
defined in analogy with Ref.~\cite{ATLAS:2018rnh} ($m_{2b}^{\rm lead(subl)}$ represent the leading (subleading) Higgs boson candidate mass from the $b\bar{b}$ pairs). Then we order these Higgs bosons according to their transverse momentum ($Hst$ and $Hnd$). 

\item We work with 15 low-level features (shown in \tabref{tab:lowf}) that characterize the final state at detector level.

\item We work with 18 high-level features as shown in \tabref{tab:highf}. They are built from the previous low-level features and from the reconstructed (reco) Higgs bosons.
\end{itemize}

\begin{table}[t!]
\begin{center}
\begin{tabular}{| c | c |} \hline
{\bf Low-level feature} & {\bf Description} \\ \hhline{|=|=|}
$N_j$ & Number of light-jets \\
$p_T^i$ & Transverse momentum of the four leading $b$-jets ($i=bst,\,bnd,\,brd,\,bth$) \\
$\eta^i$ & Pseudorapidity of the four leading $b$-jets ($i=bst,\,bnd,\,brd,\,bth$) \\
$\phi^i$ & Azimuthal angle of the four leading $b$-jets ($i=bst,\,bnd,\,brd,\,bth$) \\
$\MET$ & Missing transverse momentum \\
$\phimet$ & Azimuthal angle of the missing transverse momentum \\ \hline
\end{tabular}
\caption{List of low-level features implemented as input variables of our ML classifiers.}
\label{tab:lowf}
\end{center}
\end{table} 

\begin{table}[t!]
\begin{center}
\begin{tabular}{| c | c |} \hline
{\bf High-level feature} & {\bf Description} \\ \hhline{|=|=|}
$\chi_{hh}$ & Defined in \eref{chiHH} \\
$p_T^i$ & Transverse momentum of the two reco Higgs bosons ($i=Hst,\,Hnd$) \\
$\eta^i$ & Pseudorapidity of the two reco Higgs bosons ($i=Hst,\,Hnd$) \\
$\phi^i$ & Azimuthal angle of the two reco Higgs bosons ($i=Hst,\,Hnd$) \\
$m_{hh}$ & Invariant mass of the reco Higgs boson pair \\
$\Delta\eta_{hh}$ & Difference $\eta^{Hst}-\eta^{Hnd}$ of the two reco Higgs bosons \\ 
$\Delta\phi_{hh}$ & Difference $\phi^{Hst}-\phi^{Hnd}$ of the two reco Higgs bosons \\
$\Delta R_{hh}$ & Distance $\sqrt{\Delta\eta_{hh}^2+\Delta\phi_{hh}^2}$ of the two reco Higgs bosons \\ 
$\Delta\phi_{MET}^i$ & Differences $\phimet-\phi^i$ for $i=bst,\,bnd,\,brd,\,bth,\,Hst,\,Hnd$ \\
$\MET$ significance & Computed as $\MET/\sqrt{p_T^{bst}+p_T^{bnd}+p_T^{brd}+p_T^{bth}}$ \\ \hline
\end{tabular}
\caption{List of high-level features implemented as input variables of our ML classifiers. Here, reco means reconstructed.}
\label{tab:highf}
\end{center}
\end{table} 

We close this Section with a brief discussion of the cut-based analysis, i.e. traditional counting methods based on rectangular cuts, applied to the effective models. 
Based on the search strategies developed in~\cite{Arganda:2017wjh}, we present two cases as examples: 750\_350\_25 and 1000\_275\_25.
We consider the significance with $\Delta=15\%$ of systematic uncertainty as~\cite{Cowan2012DiscoverySF}
\begin{equation}
\label{eq:sigsys}
    \mathcal{S}_\text{sys} = \sqrt{2 \left((B+S) \log \left(\frac{\left(\Delta^2B^2+B\right) (B+S)}{\Delta^2B^2 (B+S)+B^2}\right)-\frac{1}{\Delta ^2}\log \left(\frac{\Delta^2B^2 S}{B \left(\Delta^2B^2+B\right)}+1\right)\right)} \,,
\end{equation}
where $S$ and $B$ are the number of signal and total background events at the end of the search strategy with a luminosity of $\cal{L}$ = 1 ab$^{-1}$.
Using a cut and count analysis, these two benchmarks are the only ones that exceed evidence (3$\sigma$) significance level.

For 750\_350\_25, in addition to the cuts in \eref{SRdetcuts}, we demand
\bea
&& N_j\leq 4 \,, \quad p_T^{bst} < 160 \, \text{GeV} \,, \quad p_T^{bnd} < 110 \, \text{GeV} \,, \quad p_T^{brd} < 80 \, \text{GeV}  \,, \nonumber\\
&& \chi_{hh} < 3.5 \,, \quad m_{hh} < 380 \, \text{GeV} \,, \quad \vert\Delta\phi_{MET}^{bst}\vert < 1.8 \,,
\eea
resulting in $S=22.5$ and $B=16.9$ with a significance of 3.77$\sigma$.
In addition, the corresponding distributions are collected in \fref{features-750_350_25} of \aref{App-kinem}.

On the other hand, for 1000\_275\_25, in addition to the cuts in \eref{SRdetcuts}, we demand
\be
N_j\leq 2 \,, \quad \chi_{hh} < 3 \,, \quad m_{hh} < 300 \, \text{GeV} \,, \quad \Delta R_{hh} < 1 \,,
\ee
resulting in $S=11.6$ and $B=6$ with a significance of 3.48$\sigma$.


\section{Machine-Learning Algorithms for Collider Analyses}
\label{MLalgo}

In order to tackle the binary classification problem we consider two ML algorithms, BDTs and DNNs, and study their performance in discriminating the new physics signal from the SM backgrounds. Before going into the results we provide a brief overview of the gradient-boosted trees, which is the ensemble method {\tt XGBoost} is based on, and also of neural networks. In addition, we describe the training procedure used in each case. 

\subsection{{\tt XGBoost} Overview and Architecture}
\label{MLalgo1}
The discussion in this section follows closely the {\tt XGBoost} documentation~\cite{Chen:2016btl} and Ref.~\cite{MEHTA20191}. The {\tt XGBoost} machine learning algorithm is used for supervised learning problems. There are two basic elements in supervised learning: the model and the cost or objective function. The model is the mapping between the training data $\mathbf{x}_i$, which involves multiple features, and the target variable $y_i$ to be predicted. The parameters of this mapping are learned from data by training the model, which amounts to minimizing an objective function. Once the parameters are obtained, the model provides predictions $\hat{y}_i$ of the target variable for each data point $\mathbf{x}_i$. Of course, the model can then be used to classify unlabeled inputs. The objective function contains two terms: the training loss function $l(y_i,\hat{y}_i)$, that measures how predictive the model is for each data point of the training set, and a regularization term $\Omega$ that does not depend on the data and controls the complexity of the model, which is useful to avoid overfitting.

In {\tt XGBoost} the model is an ensemble of classification and regression trees (CART). The prediction of the ensemble for a data point $(y_i,\mathbf{x}_i)$ is given by

\begin{equation}
    \label{model}
\hat{y}_i=\sum_{k=1}^K f_k(\mathbf{x}_i) \,,
\end{equation}
where $K$ is the number of trees in the ensemble and $f_k$ is a function in the functional space $\mathcal{F}$ of all possible CARTs. On the other hand, the objective function can be written as

\begin{equation}
    \label{obj}
    \mathrm{obj} = \sum_{i=1}^N l(y_i,\hat{y}_i)+\sum_{k=1}^M \Omega(f_k) \,,
\end{equation}
where $i$ runs over the data points. In order to build the ensemble in Eq.~(\ref{model}) an iterative strategy is applied by adding one new tree at a time. A family of predictors $\hat{y}^{(n)}_i$ can be defined as
\begin{equation}
    \label{prefam}
    \hat{y}^{(n)}_i =\sum_{k=1}^n f_k(\mathbf{x}_i)=\hat{y}^{(n-1)}_i+ f_n(\mathbf{x}_i) \,,
\end{equation}
and then the corresponding objective function is given by
\begin{equation}
    \label{objnth}
    \mathrm{obj}_n = \sum_{i=1}^N l(y_i, \hat{y}^{(n-1)}_i+ f_n(\mathbf{x}_i)) + \sum_{k=1}^n \Omega(f_k) \,.
\end{equation}
By assuming that the adding of a new tree is a small perturbation to the predictor, the loss function can be Taylor expanded to second order to obtain:
\begin{equation}
    \label{objit}
    \mathrm{obj}_n \approx \mathrm{obj}_{n-1} + \Delta\mathrm{obj}_n \,,
\end{equation}
with 
\begin{equation}
    \label{delobj}
    \Delta\mathrm{obj}_n = \sum^N_{i=1} g_i f_n(\mathbf{x}_i) + \frac{1}{2}h_i f_n(\mathbf{x}_i)^2 + \Omega(f_n) \,,
\end{equation}
where $g_i$ y $h_i$ are defined as
\begin{equation}
    \label{gi}
    g_i = \partial_{\hat{y}^{(n-1)}_i}l(y_i,\hat{y}^{(n-1)}_i) \,,
\end{equation}
\begin{equation}
    \label{hi}
    h_i = \partial^2_{\hat{y}^{(n-1)}_i}l(y_i,\hat{y}^{(n-1)}_i) \,.
\end{equation}
The $n$-th decision tree in the ensemble $f_n$ is then chosen to minimize $\Delta\mathrm{obj}_n$, which takes $g_i$ and $h_i$ as inputs. 

In order to specify the regularization term and to obtain an analytical expression for the parameters minimizing $\Delta\mathrm{obj}_n$, the following convenient parameterization of the decision trees is introduced:
\begin{equation}
    \label{param}
f_j(\mathbf{x}_i) = w_{q(\mathbf{x}_i)} \,,
\end{equation}
where $f_j(\mathbf{x}_i)$ is the prediction of the $j$-th tree for the data point $\mathbf{x_i}$, $q(\mathbf{x})$ is a function that assigns each data point to one of the $T$ leaves of the tree, $q:\mathbf{x}\in \mathbb{R}^d\to \{1,2,...,T\}$, and $w \in \mathbb{R}^T$ is a vector of weights on leaves. With this parameterization, the regularization term can be written as
\begin{equation}
    \label{regterm}
    \Omega(f)=\gamma T + \frac{1}{2}\lambda\sum^T_{j=1}w^2_j \,,
\end{equation}
with the parameters $\gamma$ and $\lambda$ controlling the penalty for having large partitions with many leaves and large weights on the leaves, respectively~\footnote{Notice that the second term in Eq.~(\ref{regterm}) corresponds to a $L_2$ regularization term since it depends on the $L_2$ norm of the vector of weights. A regularization in terms of the $L_1$ norm is also available in {\tt XGBoost} and is driven by a parameter called $\alpha$. }.

By using Eq.~(\ref{param}) the Eq.~(\ref{delobj}) becomes
\begin{equation}
    \label{delobj2}
    \Delta\mathrm{obj}_n = \sum^T_{j=1}[G_j w_j+\frac{1}{2}(H_j+\lambda) w^2_j] +\gamma T \,,
\end{equation}
where $G_j=\sum_{i\in I_j}g_i$ and $H_j=\sum_{i\in I_j}h_i$, with $I_j=\{i|q(\mathbf{x}_i)=j\}$ the set of indices of data points assigned to the $j$-th leaf. The optimal weights are found to be
\begin{equation}
    \label{wopt}
    w^{\mathrm{opt}}_j  =-\frac{G_j}{H_j+\lambda} \,,
\end{equation}
and the best reduction in the objective is then given by
\begin{equation}
    \label{objopt}
\Delta\mathrm{obj}^{\mathrm{opt}}_n= -\frac{1}{2}\sum^T_{j=1}\frac{G^2_j}{H_j+\lambda}+\gamma T \,.
\end{equation}
Ideally, one would list all the possible trees and select the one that minimizes $\Delta\mathrm{obj}^{\mathrm{opt}}_n$. However, this is not possible from a practical point of view. Instead, an approximate algorithm that optimizes one level of the tree at a time is applied. Specifically, optimal splits of leaves are sought by computing the corresponding gain in $\Delta\mathrm{obj}^{\mathrm{opt}}_n$. This procedure allows to obtain the structure of a tree that is a good local minimum of $\Delta\mathrm{obj}^{\mathrm{opt}}_n$ and then it is added to the ensemble. Of course, the algorithm implemented in {\tt XGBoost} goes beyond the schematic picture presented here and incorporates many additional parameters that guide the training procedure. These parameters are known as hyper-parameters since they are not learned within the estimator but need to be set externally to their optimal values. The optimization is performed by searching the hyper-parameter space for the best cross validation score.  

\begin{table}[t!]
\begin{center}
\begin{tabular}{| c | c |} \hline
{\bf Parameter} & {\bf Description} \\ \hhline{|=|=|}
{\tt learning\_rate} & Step size shrinkage of weights used at each boosting step\\
{\tt max\_depth} & Maximum depth of a tree \\
{\tt min\_child\_weight} & Minimum sum of instance weight required in a child \\
{\tt subsample} & Fraction of training instances to be random samples for each tree \\
{\tt colsample\_bytree} & Subsample ratio of columns when constructing each tree\\
{\tt gamma} & Specifies the minimum loss reduction required to split a node \\
{\tt reg\_alpha} & L1 regularization term on weights \\ 
{\tt reg\_lambda} &  L2 regularization term on weights\\ \hline
\end{tabular}
\caption{List of the hyper-parameters tuned in this {\tt XGBoost} analysis.}
\label{tab:hyperpar}
\end{center}
\end{table}

In this work, we set the objective function to {\tt binary:logistic}, which corresponds to logistic regression for binary classification and provides a probability as output. In addition, we choose the Receiver Operating Characteristic Area under the Curve (AUC) as the evaluation metric. Regarding the booster parameters, we concentrate on the optimization of those listed in Table~\ref{tab:hyperpar}, which is carried out via {\tt GridSearchCV}, a tool that exhaustively searches over a predefined parameter grid to retain the best combination. Finally, the number of boost rounds was controlled by using early stopping, a training approach that works by stopping the training procedure once the performance on a separate validation sample has not improved after a fixed number of boost rounds. 

We split the simulated sample in training, validation, and test subsamples comprising $\sim$ 65\%, 15\%, and 20\% of the total number of events, respectively. These three subsamples have a 1:1 composition of signal and background events, and the contribution of each background process to the background sample is set by following their relative cross sections after applying the selection cuts that define the signal region considered in this study. 

Several combinations of the kinematic variables presented in Section~\ref{ph-frame} have been considered as input for the training process. The best classification power was obtained by using the following set of 14 features:
\begin{eqnarray}
\label{kin-var}
&& p^{bst}_T,\eta^{bst}\,, p^{bnd}_T,\eta^{bnd}\,, p^{Hst}_T, p^{Hnd}_T\,, E^{\mathrm{miss}}_T\mbox{ sig}\,, \chi_{hh}\,, m_{hh}\,, \Delta R_{hh}\,, \nonumber \\
&& \Delta \phi^{bst}_{MET}\,, \Delta \phi^{bnd}_{MET}\,, \Delta \phi^{brd}_{MET}\,, \text{and} \, \Delta \phi^{bth}_{MET} \,.
\end{eqnarray}
In the following, we will focus exclusively on the results obtained with classifiers trained with the above set of kinematic variables.  We build a classifier for each one of the 14 benchmark models introduced in Figure~\ref{fig-effxs}.

\begin{figure}[t!]
    \begin{center}
        \includegraphics[scale=0.485]{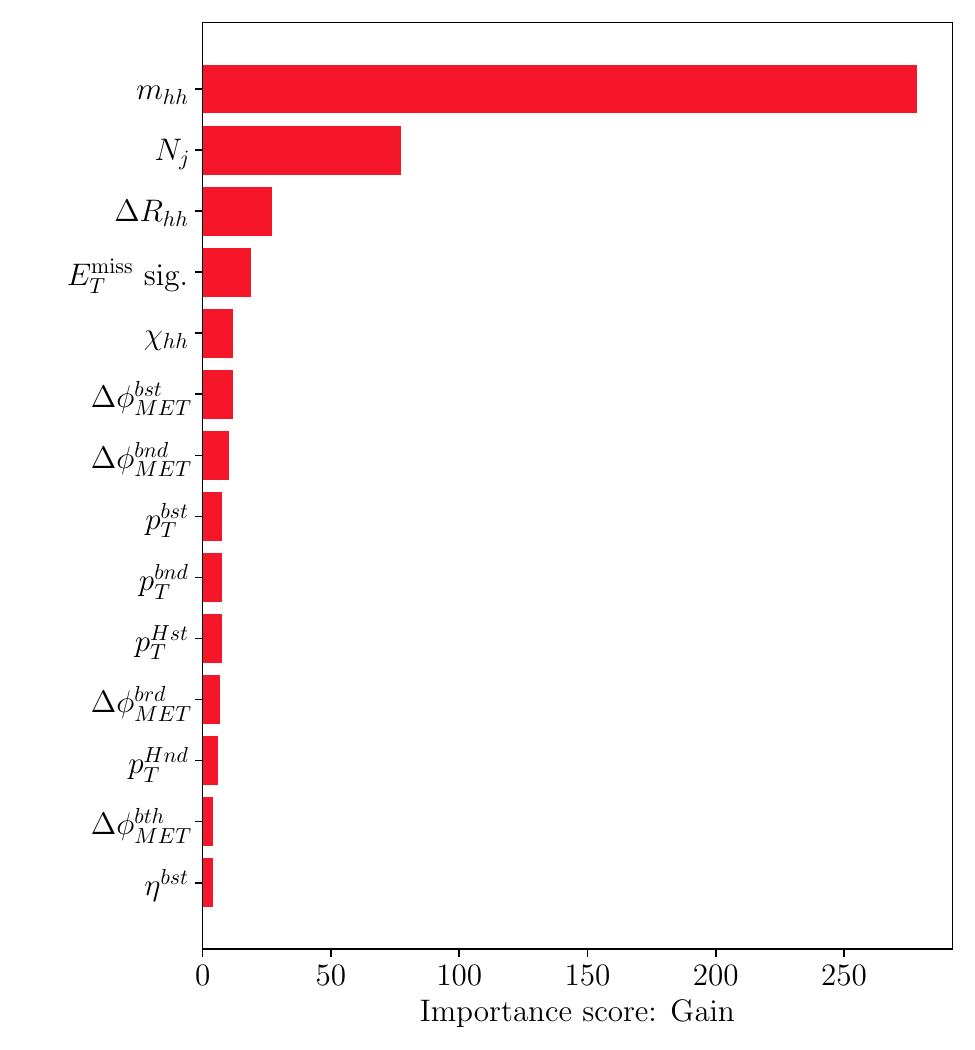}\hfill
        \includegraphics[scale=0.485]{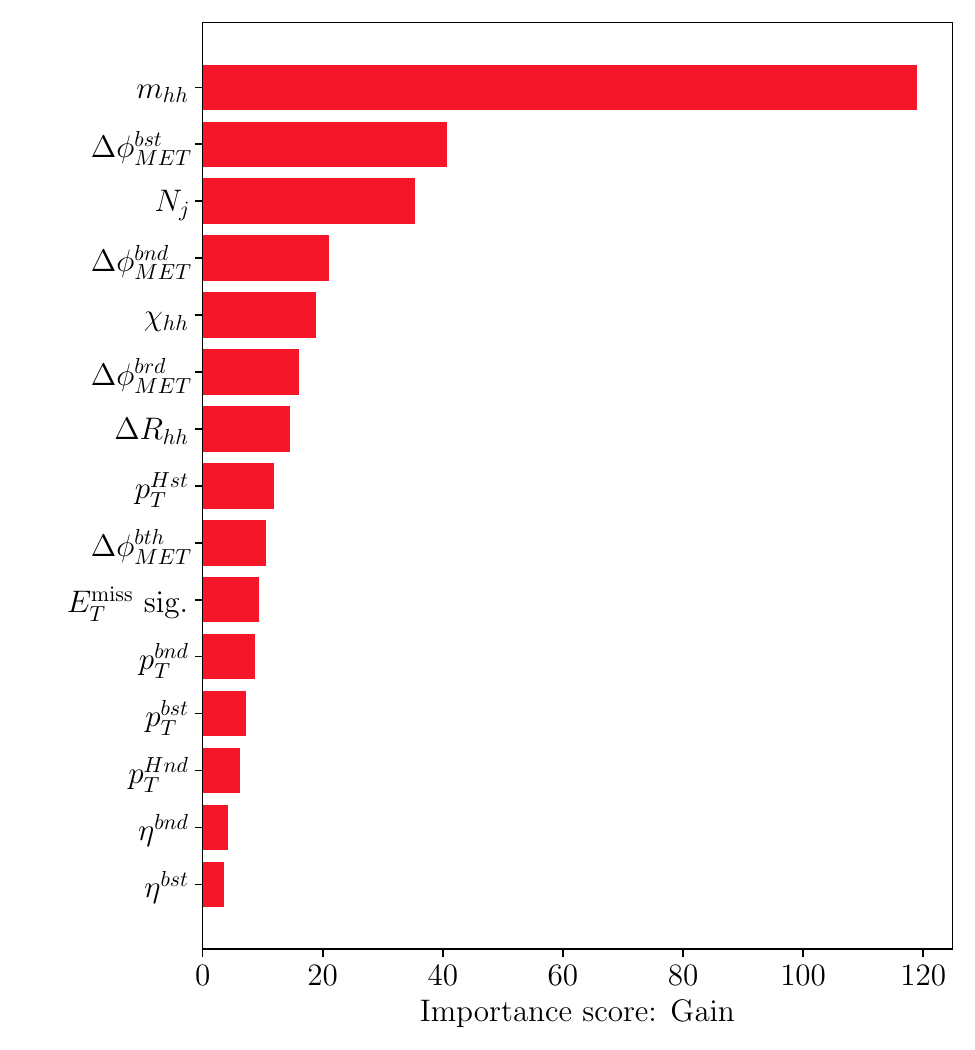} 
        \caption{Importance of the features listed in Eq.~(\ref{kin-var}) according to the gain score, for the benchmarks $750\_275\_25$ (left panel) and $750\_350\_25$ (right panel)}
        \label{fig:xgboost.imp}
    \end{center}
\end{figure}
 
In Figure~\ref{fig:xgboost.imp} we show the importance of the above features for the classifiers corresponding to two representative benchmarks in terms of the score {\tt gain}, which gives the average improvement in accuracy obtained by adding new splits on the feature. By looking at the distributions of the features (see Appendix \ref{App-kinem}) we can gain some understanding of their placing in the ranking. For example, for both benchmarks the highest ranked feature is the invariant mass $m_{hh}$, that is certainly a powerful discriminant between signal and background since it is concentrated around the intermediate scalar mass for the signal while for the backgrounds the distribution reaches it maximum at larger $m_{hh}$ values and then decreases slowly. For $750\_275\_25$ the second feature in the ranking is $N_j$, which also appears as the third variable in importance for $750\_350\_25$. Again, this is consistent with what is observed in the corresponding distributions: in both cases, the maximum is located at small $N_j$ and this allows to separate the signal from the backgrounds (except for the irreducible $Zb\bar{b}b\bar{b}$). The third most important variable for $750\_275\_25$ is $\Delta R_{hh}$, and this is consistent with the fact that the Higgs bosons tend to be boosted due to the difference in mass between the scalars $\upphi$ and $\upvarphi$, making this variable an efficient discriminator. This is not the case for $750\_350\_25$ since $\upvarphi$ is now considerably heavier, and the distributions corresponding to the signal and the different backgrounds become quite similar. In fact, we see that this variable is degraded to the 7th place for this benchmark. Finally, $\Delta \phi^{bst}_{MET}$ is the second most important feature for $750\_350\_25$ while it is in the 6th place for $750\_350\_25$, and this once again is compatible with the corresponding distributions: for $750\_350\_25$ the transverse momentum of the leading $b$-jet is more likely to lie in the same direction as the missing transverse momentum while it is in the opposite direction both for the backgrounds and the signal $750\_275\_25$; thus, it is reasonable that this feature be more useful to separate signal from background for the benchmark $750\_350\_25$.  

\subsection{DNN Overview and Architecture}
\label{DNNparton}

The discussion in this section follows closely Ref.~\cite{roberts_yaida_hanin_2022} and Keras documentation~\cite{Charles2013}. Deep learning provides an alternative way to approximate a mapping to predict the target variable $y$ based on the input vector of features $\textbf{x}$. In this case, the model is built in terms of units, known as neurons, which process a set of incoming signals $s_j$ to produce an output. Mathematically, neurons carry out two consecutive operations: first, they linearly combine the incoming signals into a preactivation variable $z$,
  \begin{equation*}
    z = b + \sum_{j=1}^{n} W_{j} \cdot s_j \,,
  \end{equation*}
where each signal is weighted by $W_{j}$ and biased by $b$; second, neurons fire or not according to the value of the preactivation variable $z$, and produce an outgoing signal
  \begin{equation*}                                                                 
    \sigma = \hat{\sigma}(z) \,.
  \end{equation*} 
The scalar-valued function $\hat{\sigma}(z)$ is called the activation function and is often taken as a rectified linear unit (ReLU) function to introduce a controllable degree of non-linearity in the model.

A set of neurons is called a layer, and stacked layers, inter-connected in such a way output signals from one layer work as the input of those neurons that belong to the subsequent layer, constitute DNN. Iteratively, DNNs are defined as
\begin{align*}
	z_i^{(1)} &= b_i^{(1)} + \sum_{j=1}^{n_1} W_{ij}^{(1)} \cdot s_j \,,
	\\
	& \vdots
	\\
	z_i^{(\ell-1)} &= b_j^{(\ell-1)} + \sum_{j=1}^{n_{\ell-1}} W_{ij}^{(\ell-1)} \cdot \hat\sigma\big(z_j^{(\ell-2)}\big) \,,
	\\
	z_i^{(\ell)} &= b_i^{(\ell)} + \sum_{j=1}^{n_\ell} W_{ij}^{(\ell)} \cdot \hat\sigma\big(z_j^{(\ell-1)}\big) \,.
\end{align*}
The number of layers $\ell$ (depth), the number of neurons per layer $n_\ell$, and the patterns of layer connections determine the network architecture. 

DNNs for binary classification tasks typically count with a single neuron in the outgoing layer, activated by a sigmoid function, instead of a ReLU, considering the DNN output is interpreted as probability. In our case, we based the architecture of the classification model on the work done in Ref.~\cite{Alvestad:2021sje} for sneutrino searches, setting $500$, $500$, $250$, $100$, and $50$ neurons in five consecutive layers. To endow the network architecture with a certain level of flexibility and reduce the risk of overfitting, we set up a probability of discarding outputs of intermediate layers or dropout-rate of $21\%$.

Similarly to what was explained in the previous section, a DNN model provides a prediction $\hat{y}_i$ of the target variable $y_i$ for every data point $\textbf{x}_i$ in the training data set. Target and prediction are compared through an objective function, defined in terms of the training loss function $l$ as
\begin{equation*}
  \text{obj} = \sum_{i=1}^N l(\hat{y}_i,y_i) \,.
\end{equation*}
In this case, we took the binary cross-entropy for $l$, which is the typical choice for binary classification problems. To optimize the model, we implemented an adaptative stochastic gradient algorithm to minimize the objective function. Schematically, this kind of algorithm updates the free parameters as follows
\begin{equation*}
  \theta_\lambda^{(n+1)} =  \theta_\lambda^{(n)} - \eta \frac{\partial \, \text{obj}}{\partial \theta_\lambda} {\bigg|}_{\theta = \theta^{(n)}} \,,
\end{equation*}
where $n$ indicates the iteration of the optimization. The learning rate $\eta$ controls the size of the step taken in the parameter space and is estimated in every iteration to improve the convergence speed. We opted for using  Adam algorithm~\cite{kingma2014adam} with an initial learning rate value of 0.001. As in the case of {\tt XGBoost}, we chose the AUC as a metric to evaluate the training performance and used the early stopping approach to avoid overtraining the model.

 \begin{figure}[t!]
    \begin{center}
        \includegraphics[scale=0.485]{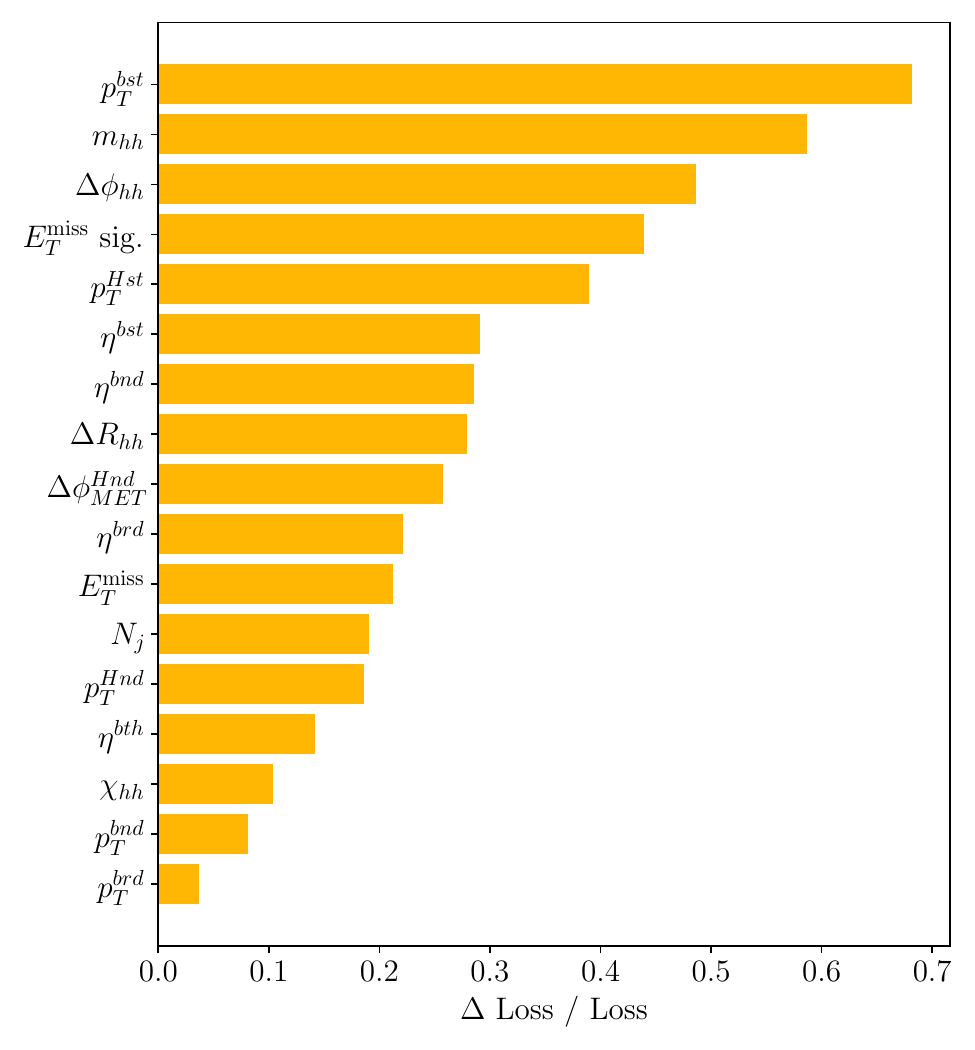}\hfill
        \includegraphics[scale=0.485]{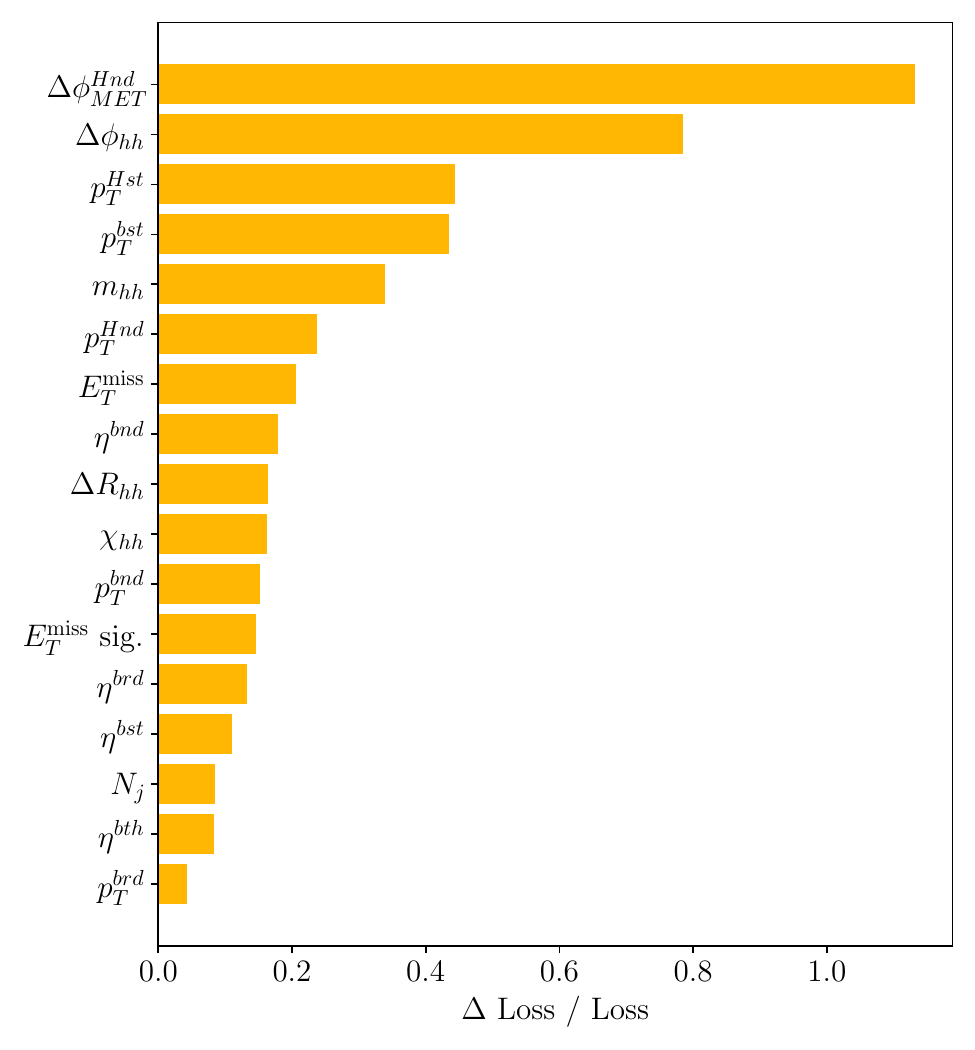} 
        \caption{Ranking of the features listed in Eq.~(\ref{eq:dnn_feat_list}) according to the permutation importance score, for the benchmarks $750\_275\_25$ (left panel) and $750\_350\_25$ (right panel)}
        \label{fig:dnn.imp}
    \end{center}
 \end{figure}

The sample used for training the DNN model includes the same events used for {\tt XGBoost} split into training, validation, and test subsamples containing also $\sim$ 65\%, 15\%, and 20\% of all the events, respectively. We tested several combinations of the kinematic variables in Section~\ref{ph-frame} as input for the training process. The best classification resulted from training the DNN model with the following set of 17 features:

\begin{align}
    &N_j\,, p_T^{bst}\,, \eta^{bst}\,, p_T^{bnd}\,, \eta^{bnd}\,, p_T^{brd}\,, \eta^{brd} \,, \eta^{bth}\,, p_T^{Hst}\,, p_T^{Hnd}\,, E_T^\text{miss}\,, E_T^\text{miss}\text{ sign.}\,, \chi_{hh}\,, m_{hh}\,, \nonumber \\
    &\Delta R_{hh}\,, \Delta\phi_{hh}\,, \text{and} \, \Delta\phi^{Hnd}_{MET}\,.
    \label{eq:dnn_feat_list}
\end{align}
All features were rescaled in order to be processed by the DNN model: the method of standardization, included among {\tt Scikit-learn} tools\cite{scikit-learn}, gave the best discrimination results. In the subsequent discussion, we will entirely focus on the results achieved with classifiers trained using the set of kinematic variables mentioned above.

In Figure \ref{fig:dnn.imp} we show two importance rankings of the features processed by two DNN models, trained to differentiate between benchmarks $750\_275\_25$ and $750\_350\_25$ events, respectively, from the background. Both rankings were performed using the permutation importance method, which scores every feature by breaking its relationship to the target and evaluating the predictive power deterioration of the model. Although DNN non-linear combinations of the input features used to differentiate between classification classes are not easily accessible, scrutinizing kinematic distributions related to these features helps to achieve some level of interpretability of the classification model. For example, for both benchmarks the signal distributions of $p_T^{bst}$ and $p_T^{Hst}$ peak in a different region than the corresponding background distributions, see \aref{App-kinem}. This property makes them suitable to discriminate between these two classes. Another example is the case $m_{hh}$: the signal distributions are concentrated around the value of $2m_h$, unlikely the background which has a much flatter distribution. As expected, these three variables rank among the highest-scored features.

\section{Results}
\label{res}

\begin{figure}[t!]
	\begin{center}
		\begin{tabular}{cc}
			\centering
			\hspace*{-16mm}
			\includegraphics[scale=0.55]{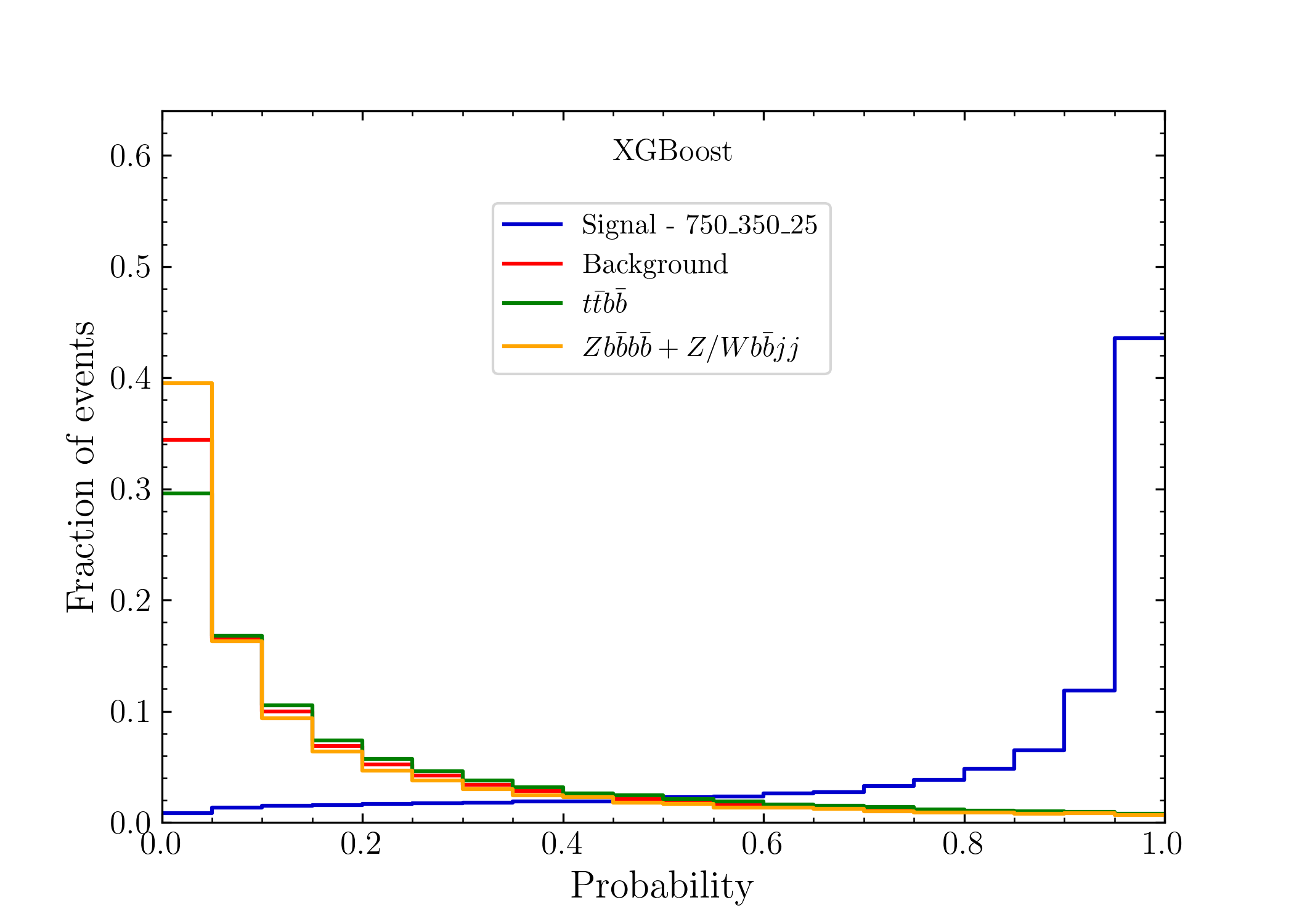} &
			\hspace{-10mm}
			\includegraphics[scale=0.55]{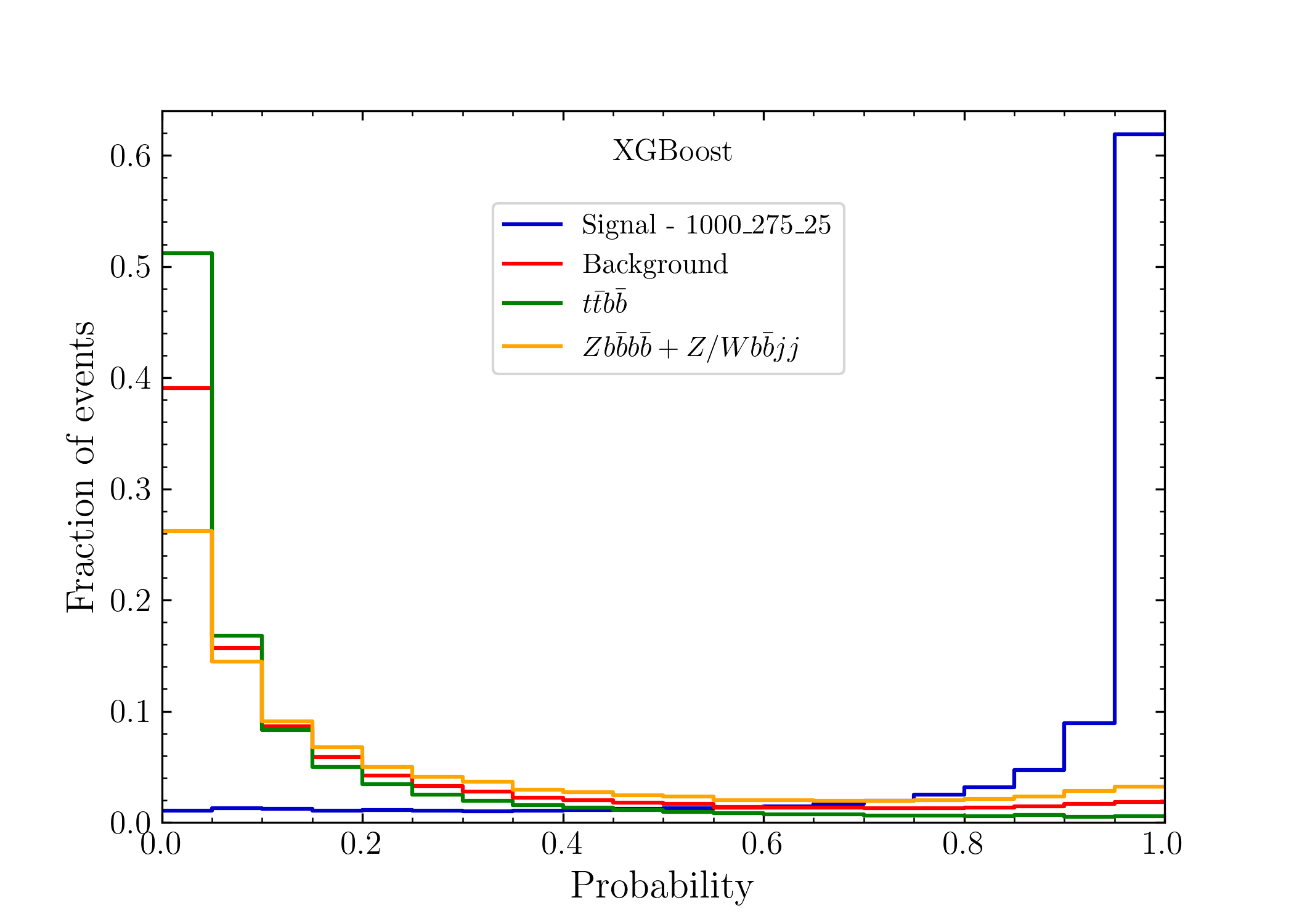} \\
            \hspace*{-16mm}
               \includegraphics[scale=0.55]{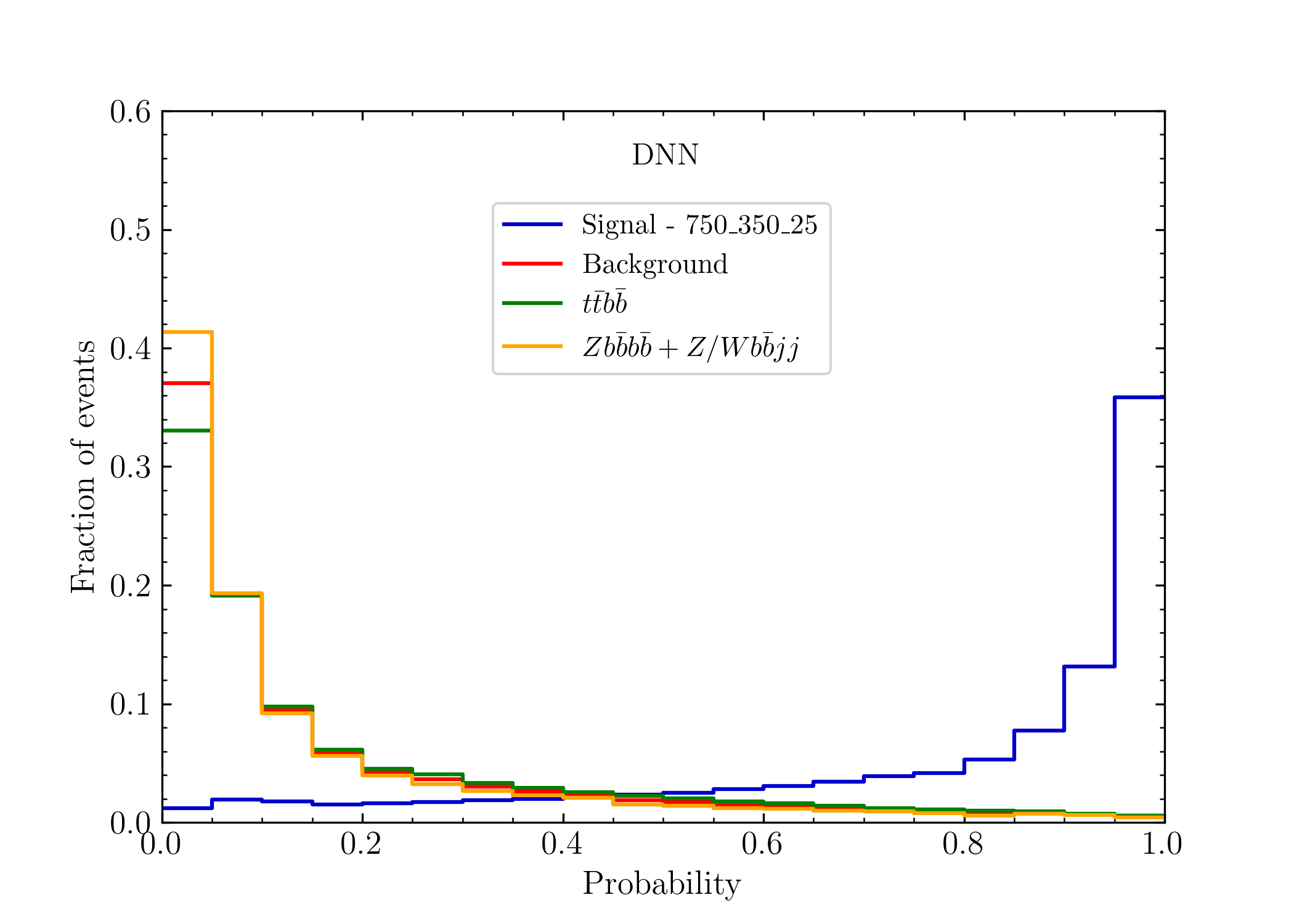} & \hspace{-10mm}
			\includegraphics[scale=0.55]{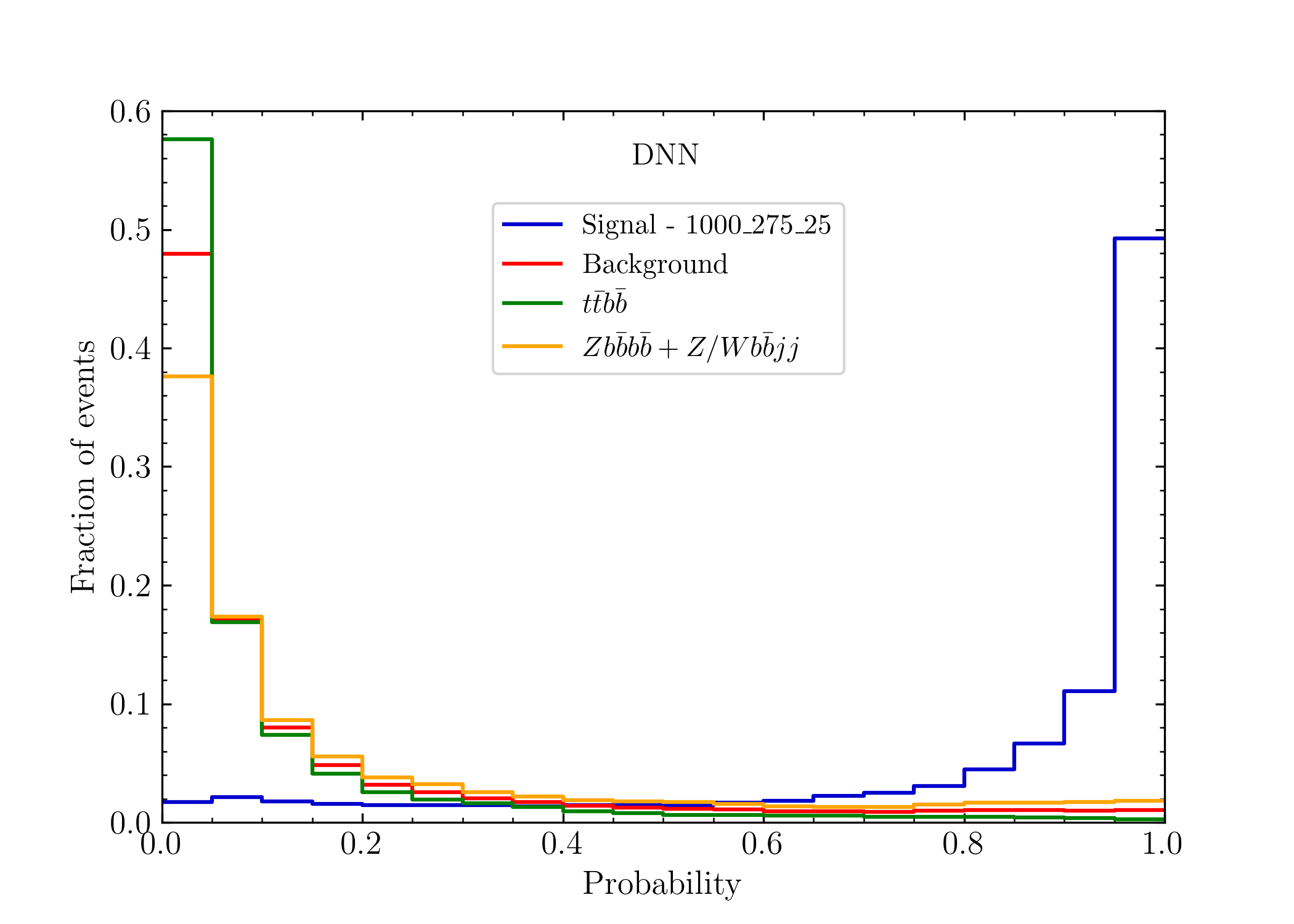}
		\end{tabular}
		\caption{Probability distributions given by the output of the {\tt XGBoost} and DNN classifiers for benchmarks $750\_350\_25$ and $1000\_275\_25$ along with the dominant backgrounds.}
		\label{fig:prob_distr}
	\end{center}
\end{figure}

For each one of the benchmarks, we apply the corresponding {\tt XGBoost} and DNN classifiers to the test samples. The output is the probability of an event being classified as a signal event. As an example, we show in Figure~\ref{fig:prob_distr} the probability distributions for the models $750\_350\_25$ and $1000\_275\_25$. We see that both ML algorithms lead to an efficient classification of signal and background events. This is not exclusive of the two displayed benchmarks since in all the cases AUC values above 0.9 are obtained. For $1000\_275\_25$, both classifiers are better at identifying $t\bar{t}b\bar{b}$ than the rest of the background processes and this is the case for most of the benchmarks (10 of 14), the model  $750\_350\_25$ is actually one of the four exceptions. Another interesting observation is that while the {\tt XGBoost} classifier seems to label the signal events better, its DNN counterpart classifies the background events more efficiently. Of course, the precise values for the signal acceptance and background rejection will depend on the chosen probability threshold.  

\begin{figure}[t!]
    \begin{center}
    \begin{tabular}{cc}
    \centering
    \hspace*{-10mm}
        \includegraphics[width=.52\textwidth]{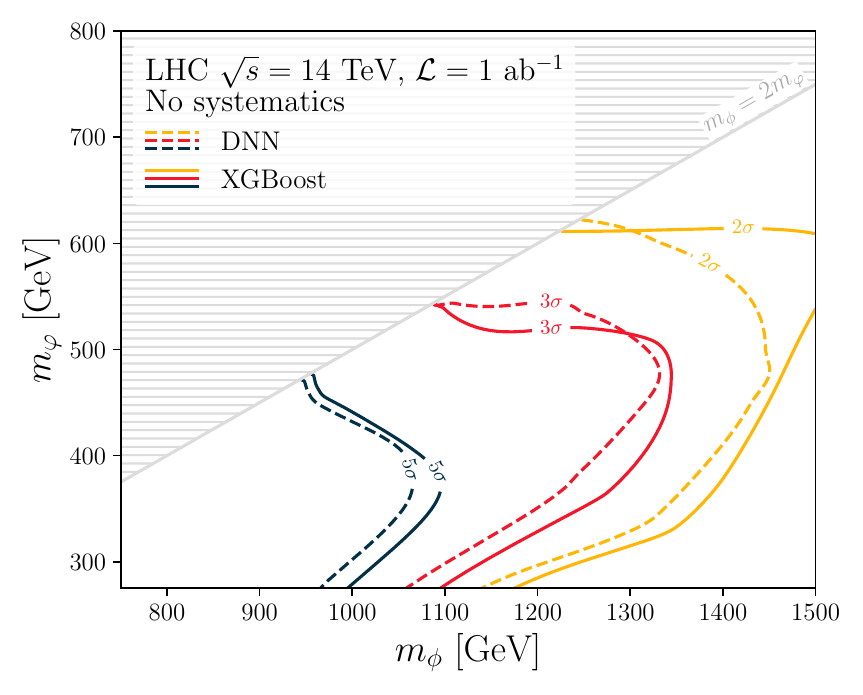}
        & \includegraphics[width=.52\textwidth]{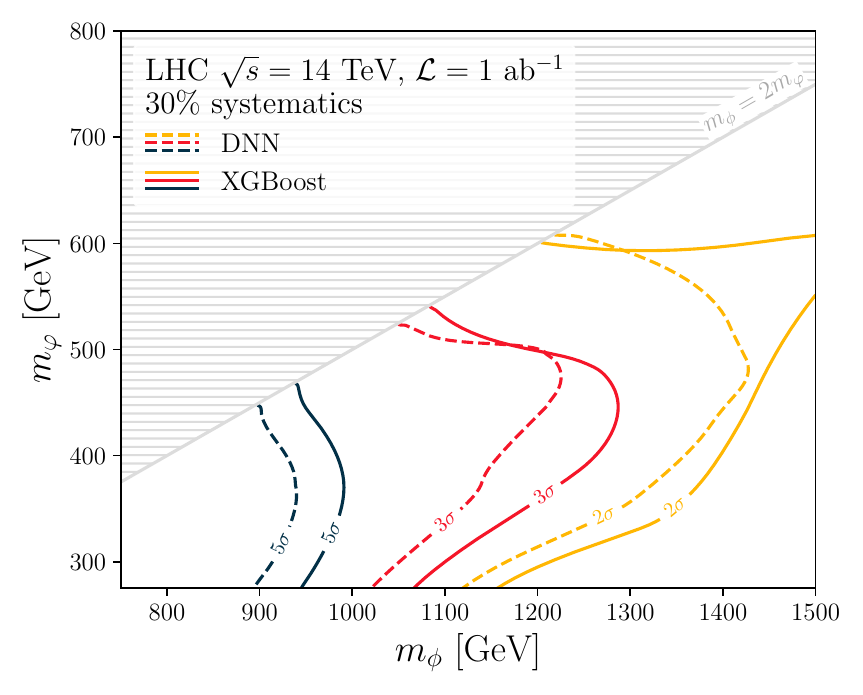}
        \end{tabular}
        \caption{Significance level curves for {\tt XGBoost} and DNN algorithms in the plane $[m_\phi,m_\upvarphi]$ with no (left panel) and 30\% (right panel) systematics. The hatched region for $m_\phi < 2m_\upvarphi$ is kinematically forbbiden.}
        \label{fig:results}
    \end{center}
\end{figure}

For each benchmark, the probability threshold is chosen to maximize the significance computed as~\cite{Cowan:2010js}
\begin{equation}
    \label{eq:sig}
    \mathcal{S}=\sqrt{-2\left((S+B)\ln\left(\frac{B}{S+B}\right)+S\right)},
\end{equation}
when potential systematic uncertainties are neglected, and using Eq.~(\ref{eq:sigsys}) otherwise. In the above expression, $S$ and $B$ are the signal and background rates after applying the cuts in Eq.~(\ref{SRdetcuts}) and the additional cut on the output of the classifiers. By interpolating the results of the 14 simulated benchmarks, we are able to plot contours of constant significance in the plane $m_{\upvarphi}$-$m_{\upphi}$. The results for {\tt XGBoost} and DNNs are shown in Figure~\ref{fig:results}, where we display contours corresponding to 2$\sigma$, 3$\sigma$, and 5$\sigma$ for a luminosity of 1 ab$^{-1}$. Since the performance of the classifiers of each benchmark are quite similar, with AUC above 0.9 in all the cases, the shape of the contours is mostly driven by the effective cross section. In fact, by comparing with Figure~\ref{fig-effxs}, it is easy to recognize the behavior of the effective cross section in the contours displayed in Figure~\ref{fig:results}.

With systematic uncertainties neglected, we see that for $m_{\upphi}$ below $\sim 950$ GeV discovery-level significances (${\cal S}$ = 5$\sigma$) are reached for all the kinematically allowed intermediate scalar $\upvarphi$ masses. Moreover, for $m_{\upvarphi}$ around 380 GeV, the discovery region extends in the $m_{\upphi}$-range up to $\sim 1090$ GeV with {\tt XGBoost} and $\sim 1060$ GeV with DNNs. Evidence-level significances (${\cal S}$ = 3$\sigma$) can be obtained for $\upphi$ as heavy as $\sim 1340$ GeV providing the intermediate scalar $\upvarphi$ mass is $\sim 480$ GeV. In general, when the impact of the systematic uncertainties is not included, the results of ${\tt XGBoost}$ tend to slightly improve those of the DNNs.

In the right panel of Figure~\ref{fig:results} we show how the results are degraded when 30\% of systematic uncertainties on the total background are included. Since the expression in Eq.~(\ref{eq:sigsys}) penalizes the presence of background, now the thresholds optimizing the significance shift to increase the background rejection at the cost of decreasing the signal acceptance. This can be directly verified by comparing the acceptances of signal and backgrounds obtained before and after including the systematic uncertainties (see Tables~\ref{tab:acdnn}-\ref{tab:acxgb30} in the Appendix \ref{App-signi}). Discovery-level significances are obtained for all the kinematically allowed range of $m_{\upvarphi}$ for $m_{\upphi}<900$ GeV, and the discovery region reaches heavy scalar masses of $\sim 990$ GeV for $m_{\upvarphi}$ around 370 GeV. The evidence level can be reached now for $m_{\upphi}$ up to $\sim 1290$ GeV when $m_{\upvarphi}\sim 450$ GeV. Again, the  {\tt XGBoost} classifiers appear to provide better prospects. Moreover, the improvement is more significant when the impact of systematic uncertainties is included. This is consistent with the acceptances displayed in Tables ~\ref{tab:acdnn}-\ref{tab:acxgb30} of Appendix \ref{App-signi}, where we see that DNN classifiers keep more signal events than {\tt XGBoost} classifiers, but at the expense of rejecting considerably less background events. The improvement in signal acceptance is not enough to compensate for the worse background rejection, which is strongly penalized by Eq.~(\ref{eq:sigsys}). The fact that after applying the probability threshold the DNN classifiers keep more background events is not in conflict with the observation made before about their better discrimination of the background. Since the background is better classified than the signal, a threshold that keeps as many signal events as possible is preferred, even when this means keeping more background events. In contrast, the {\tt XGBoost} classifiers are better at classifying the signal, and then a threshold that rejects as many background events as possible is optimal, even when it leads to a smaller signal acceptance. 

The results in Figure~\ref{fig:results} show a significant improvement with respect to the prospects achieved by using a traditional cut and count analysis. In particular, when a 15\% of systematics uncertainties is considered, significances above the evidence level are obtained for 7 ({\tt XGBoost}) and 6 (DNN) out of the 14 simulated benchmarks, while with the analysis based on rectangular cuts only two benchmarks, 750\_350\_25 and 1000\_275\_25, lead to significances above $3\sigma$. As pointed out in Section~\ref{ph-frame}, the significances reached with the cut and count analysis for these benchmarks are $3.77$ and $3.48$, respectively. The performance of the ML algorithms is significantly better, giving rise to 11.75 ({\tt XGBoost}) and 10.67 (DNN) for  750\_350\_25, and 4.51 ({\tt XGBoost}) and 3.78 (DNN) for 1000\_275\_25. Moreover, if the potential systematic uncertainties on the total background are increased to 30\% , the significances obtained with the cut and count analysis drop below the evidence level, while this is not the case when the ML classifiers are used. 

\section{Conclusions}
\label{conclu}

In this paper we study the performance of two types of modern machine-learning algorithms, namely {\tt XGBoost} and deep neural networks, on the LHC signature consisting of 4 $b$-jets and large missing transverse energy, comparing it against traditional analyses based on rectangular cuts. We work within the context of simplified models that generically parameterize a large class of models with heavy scalars and dark matter candidates, consisting of an extended scalar sector with three real scalar particles: the heaviest of these new scalars, $\upphi$, produced via gluon fusion at the LHC, predominantly decays to a pair of intermediate scalars $\upvarphi$, which interact with the visible sector only through its coupling with the SM Higgs boson $h$; the third and lightest scalar $\chi$ is the DM candidate within this effective field theory framework. Therefore, the proposed LHC signature comes from the resonant production of $\upphi$, that decays into a pair of $\upvarphi$. One of these $\upvarphi$ decays in turn into two $h$, decaying both into $b$-quarks pairs, while the other $\upvarphi$ decays invisibly into a pair of $\chi$. 
The dominant irreducible backgrounds are $Z+b\bar{b}b\bar{b}$ and $t\bar{t}+b\bar{b}$; the main reducible backgrounds correspond to $V$+jets (concretely $Z/W+b\bar{b}+jj$), while QCD multijet can be safely ignored since it does not provide a true source of $E_T^\text{miss}$.

We scan $m_\upphi$ and $m_\upvarphi$ in the ranges [750, 1500] GeV and [275, $m_\upphi/2$] GeV, respectively, and consider 15 low-level and 18 high-level kinematic features with which we feed the ML algorithms. The discriminating power of these detector-level features varies by benchmark, but we have found that in general the most important are those related to the $hh$ system and $E_T^\text{miss}$, along with the $p_T$ of the most energetic jets.

Our performance comparison between both ML algorithms is based on the maximum significance reached for an LHC center-of-mass energy of 14 TeV and a total integrated luminosity of 1 ab$^{-1}$. Both algorithms present very similar performances and a significant improvement with respect to the prospects achieved by using a traditional cut and count analysis. 
In most of the parameter space, the results of ${\tt XGBoost}$ tend to slightly improve those of the DNNs due to the signal acceptance and background rejection interplay in each case when the significance is maximized.
If a 15\% of systematics uncertainties on the background is considered, significances larger than the evidence level are obtained for a large part of the 14 simulated benchmarks, while with the cut-based analysis only two benchmarks, 750\_350\_25 and 1000\_275\_25, provide significances above $3\sigma$ ($3.77$ and $3.48$, respectively). The signal significances from {\tt XGBoost} and DNN are much better: 11.75 ({\tt XGBoost}) and 10.67 (DNN) for  750\_350\_25, and 4.51 ({\tt XGBoost}) and 3.78 (DNN) for 1000\_275\_25. In addition, if a 30\% systematic uncertainty on the total background is considered, the significances obtained with the traditional analysis drop below the evidence level, whilst the ML classifiers still provide values up to the discovery level.

As a general conclusion, we consider that our phenomenological analysis shows that the proposed LHC signature deserves the development of dedicated searches by the experimental collaborations, for which modern ML algorithms such as {\tt XGBoost} and DNN would play a crucial role. 

\section*{Acknowledgments}
EA acknowledges partial financial support by the ``Atracci\'on de Talento'' program (Modalidad 1) of the Comunidad de Madrid (Spain) under the grant number 2019-T1/TIC-14019, and by the Spanish Research Agency (Agencia Estatal de Investigaci\'on) through the Grants IFT Centro de Excelencia Severo Ochoa No CEX2020-001007-S and PID2021-124704NB-I00 funded by MCIN/AEI/10.13039/ 501100011033.
The work of RAM has received financial support from CONICET and ANPCyT under projects PICT 2018-03682 and PICT-2021-00374. 

\section*{Data Availability Statement}

No Data associated in the manuscript.

\section*{Appendices}
\appendix
\section{Cross Sections}
\label{App-xs}
In Tables~\ref{tab:effxs-sig} and~\ref{tab:effxs-bkg} we provide the effective cross sections for the signal benchmarks and the backgrounds, respectively. The effective cross section is defined as $\sigma_{\mathrm{eff}}=\sigma \cdot \epsilon_{\mathrm{SR}}$, with $\sigma$ the cross section of the process and $\epsilon_{\mathrm{SR}}$ the fraction of events passing the selection cuts given in Eq.~(\ref{SRdetcuts}).
\newpage
\begin{table}[h!]
    \footnotesize
    \centering
    \begin{tabular}{|c|c|c|}
        \hline
        $m_\upphi$& $m_\upvarphi$& $\sigma_{\mathrm{eff}}\,[\mathrm{fb}]$\\ 
        \hline
         750&   275&   0.21\\
         750&   350&   0.19\\
        1000&   275&   0.044\\
        1000&   375&   0.098\\
        1000&   475&   0.036\\
        1250&   275&   0.0094\\
        1250&   375&   0.024\\
        1250&   475&   0.030\\
        1250&   600&   0.013\\
        1500&   275&   0.0013\\
        1500&   375&   0.0046\\
        1500&   475&   0.0085\\
        1500&   600&   0.0087\\
        1500&   725&   0.0037\\
        \hline
    \end{tabular}
    \caption{Effective cross section for the benchmarks considered in this work.}
    \label{tab:effxs-sig}
\end{table}
\begin{table}[h!]
    \footnotesize
    \centering
    \begin{tabular}{|c|c|}
        \hline
        Process & $\sigma_{\mathrm{eff}}\,[\mathrm{fb}]$\\ 
        \hline
        $Z\vert_{\rm inv}b\bar{b}jj$ & 0.937\\
        $\ttbbsemilep$ &   0.680\\
        $\ttbbsemitau$ &  0.580\\
        $W^\pm\vert_{\rm semilep}b\bar{b}jj$ &   0.242\\
        $\zbbbb$ & 0.112\\
            \hline
    \end{tabular}
    \caption{Effective cross section for the backgrounds considered in this work.}
    \label{tab:effxs-bkg}
\end{table}

\section{Relevant Kinematic Distributions}
\label{App-kinem}
\begin{figure}[!htbp]
    \centering
    \includegraphics[width=0.4\textwidth]{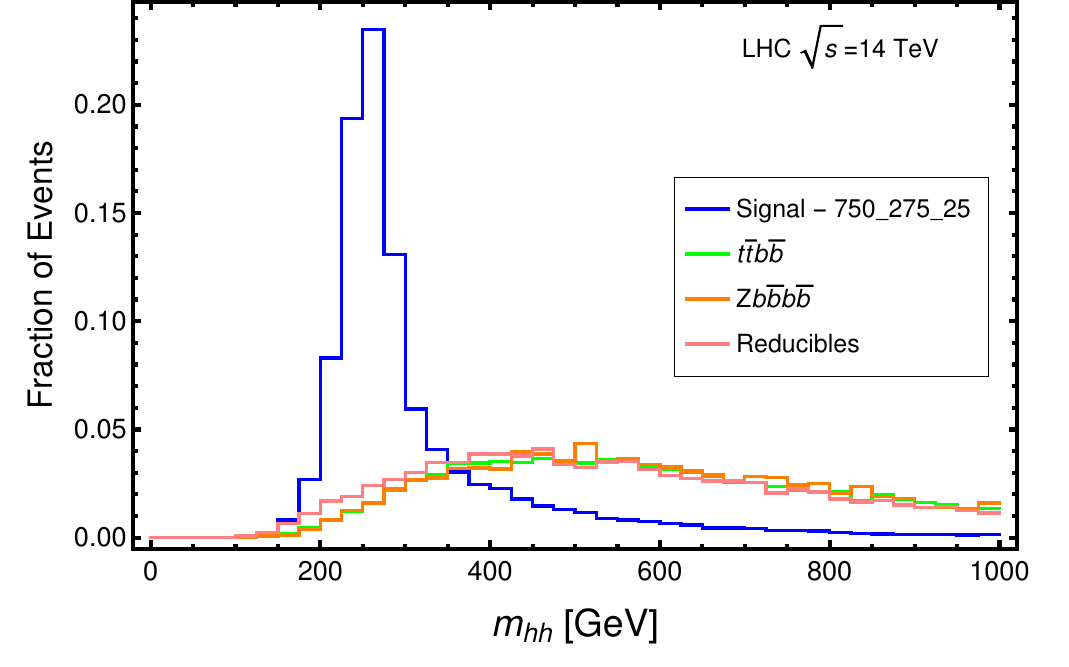}
\begin{tabular}{cc}
\includegraphics[width=0.4\textwidth]{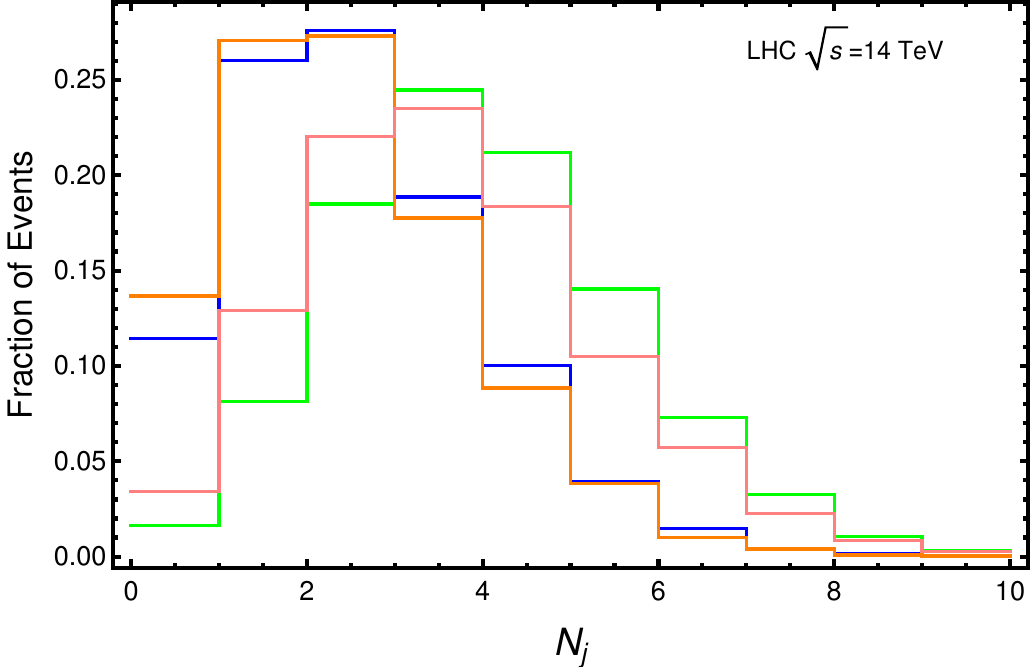} & \includegraphics[width=0.4\textwidth]{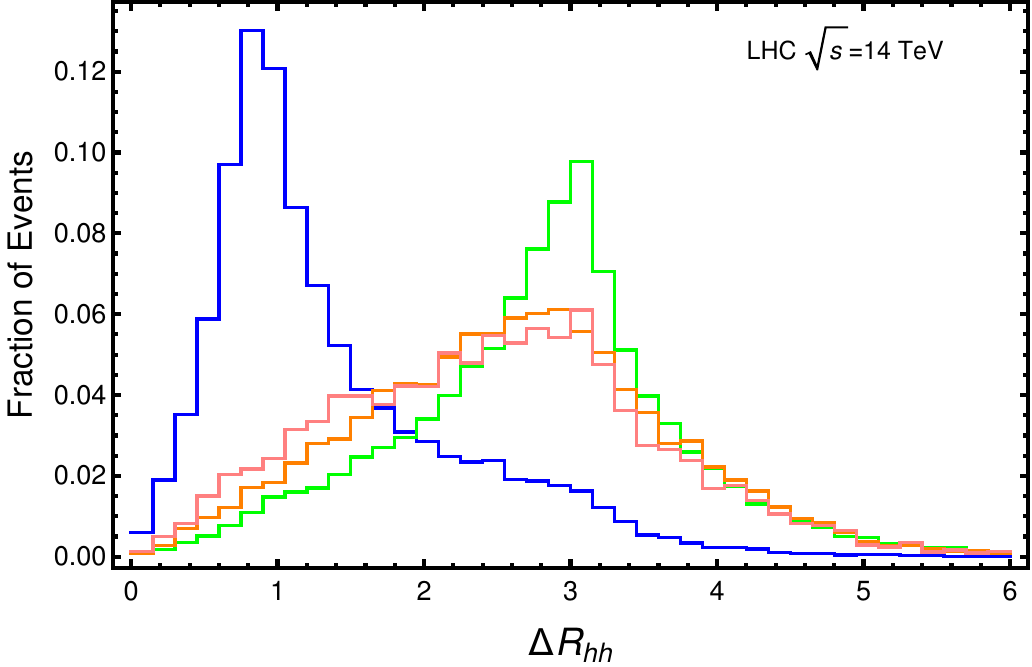} \\
\includegraphics[width=0.4\textwidth]{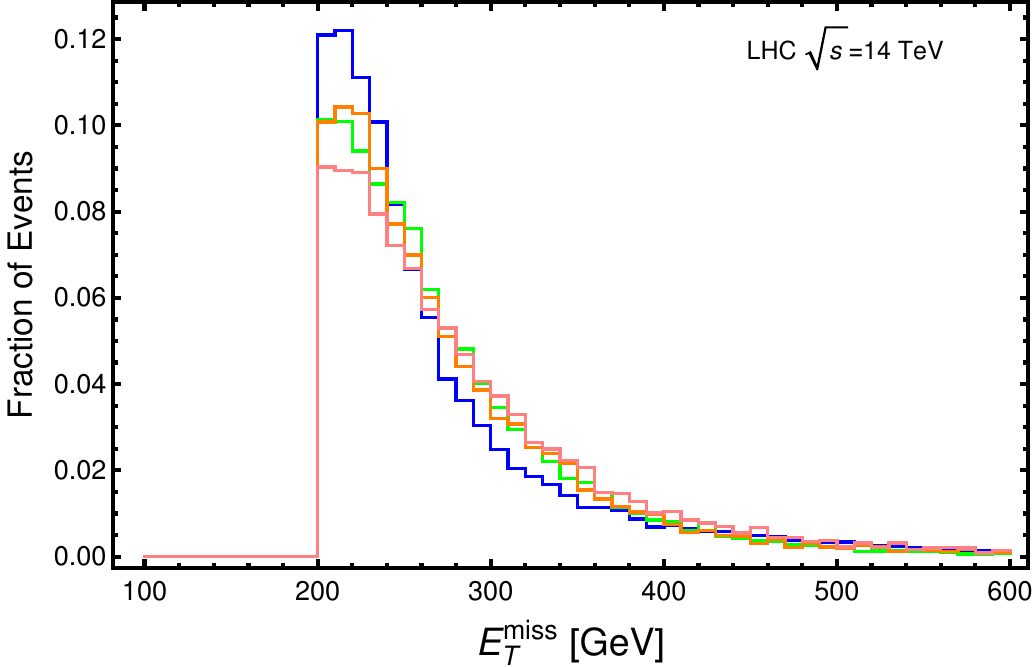} & \includegraphics[width=0.4\textwidth]{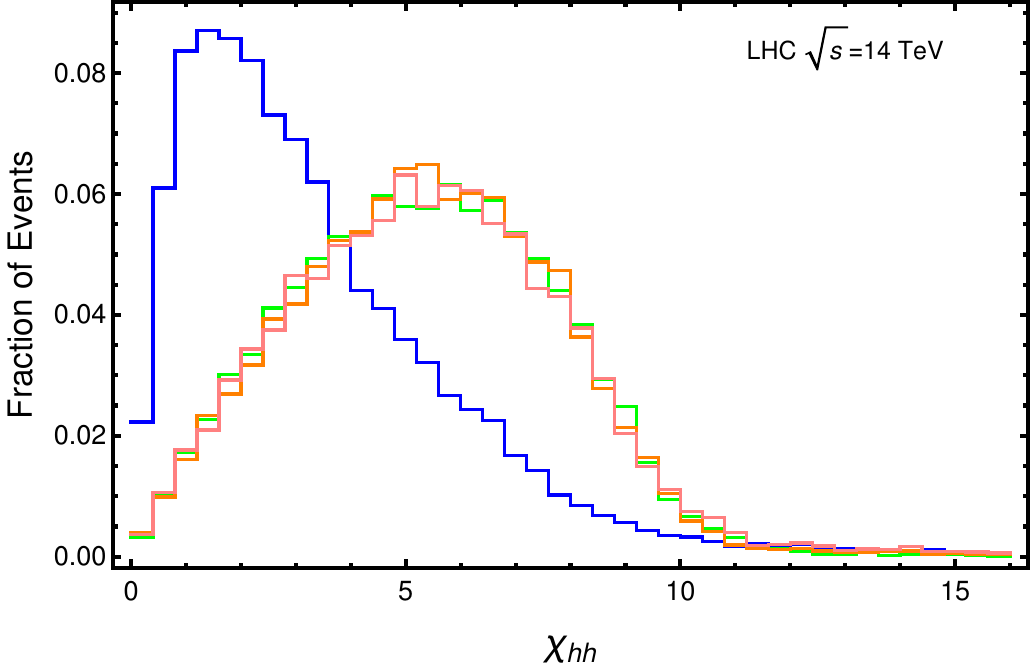} \\
\includegraphics[width=0.4\textwidth]{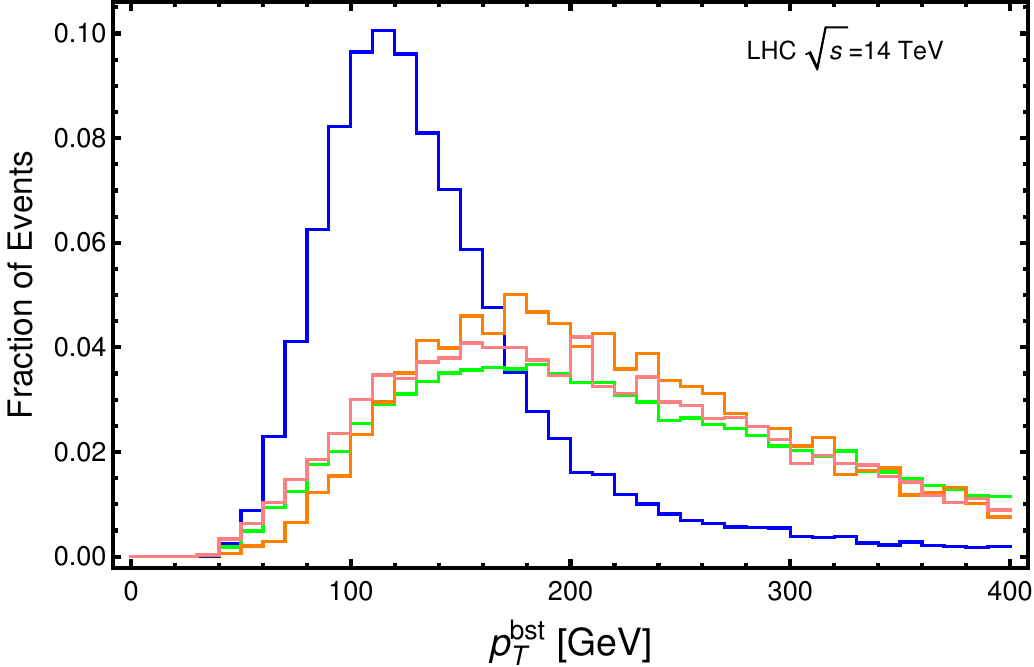} & \includegraphics[width=0.4\textwidth]{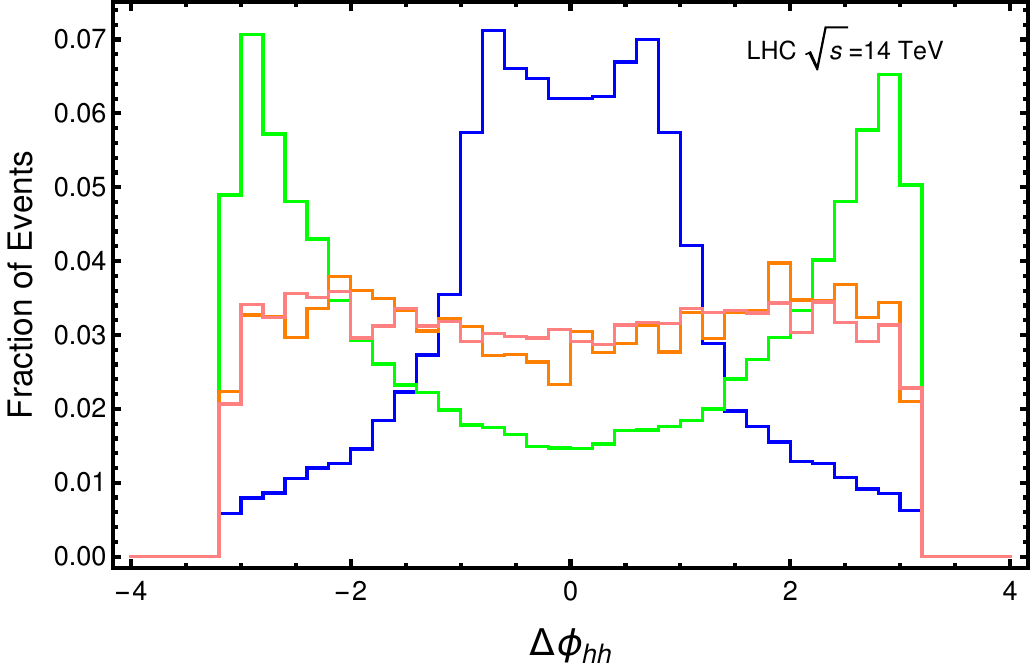} \\
\includegraphics[width=0.4\textwidth]{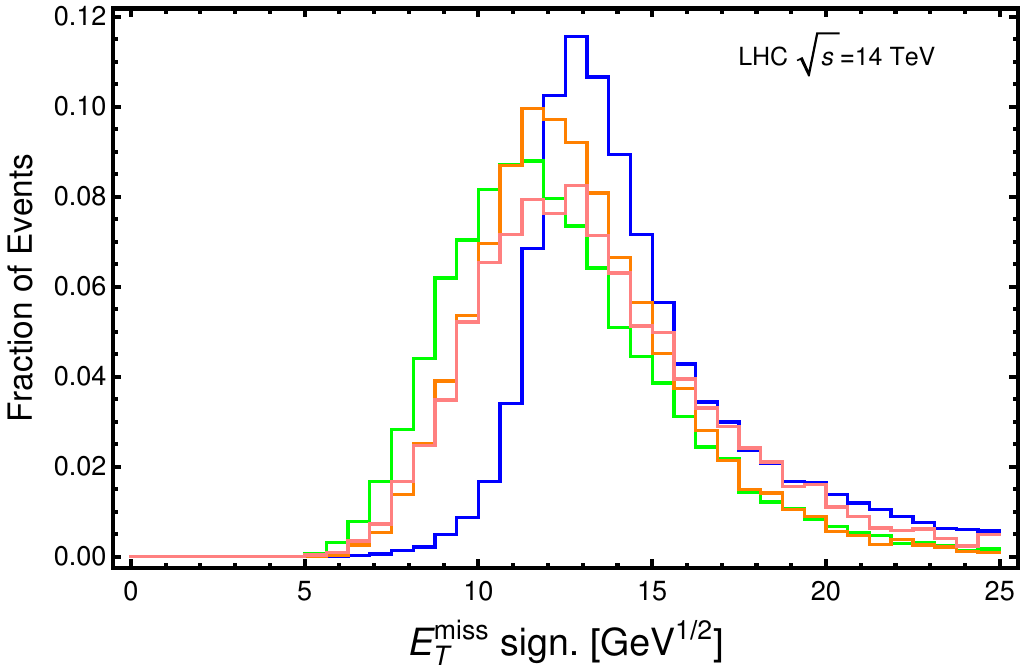} & \includegraphics[width=0.4\textwidth]{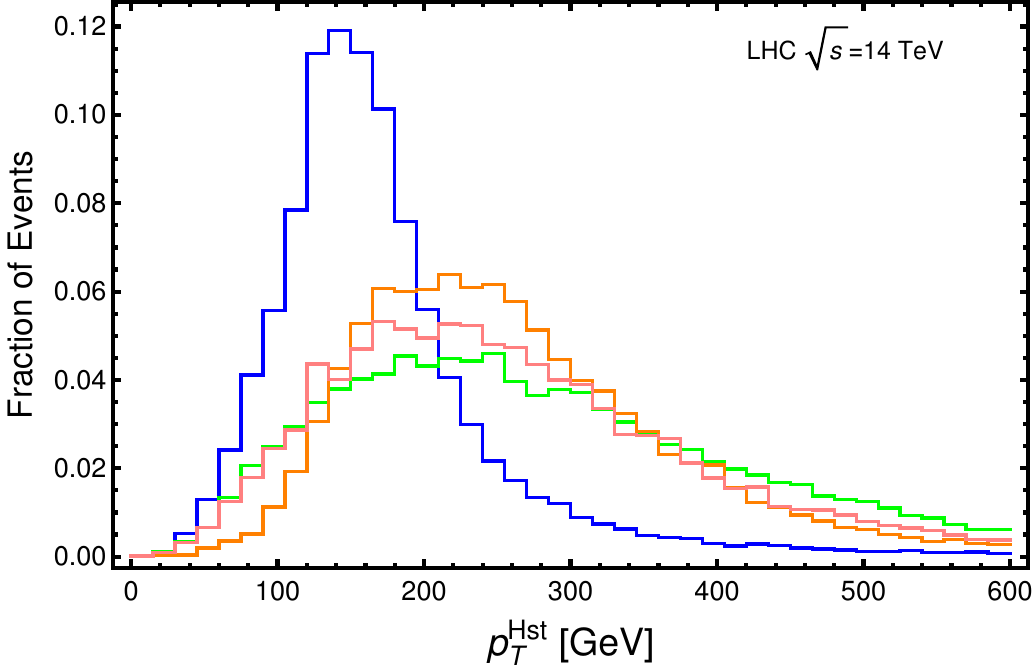} \\
\end{tabular}
\caption{Distributions of some of the kinematic variables used as inputs during the training of {\tt XGBoost} and DNNs classifiers for the benchmark 750\_275\_25.}
\label{features-750_275_25}
\end{figure}

\begin{figure}[!htbp]
    \centering
    \includegraphics[width=0.42\textwidth]{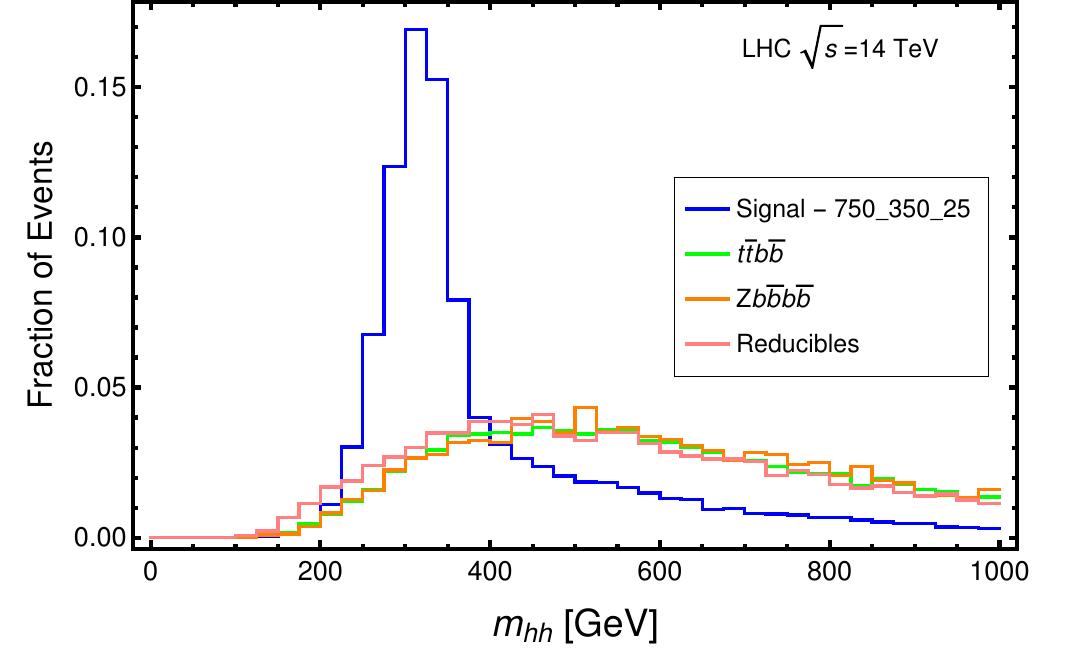}
\begin{tabular}{cc}
\includegraphics[width=0.42\textwidth]{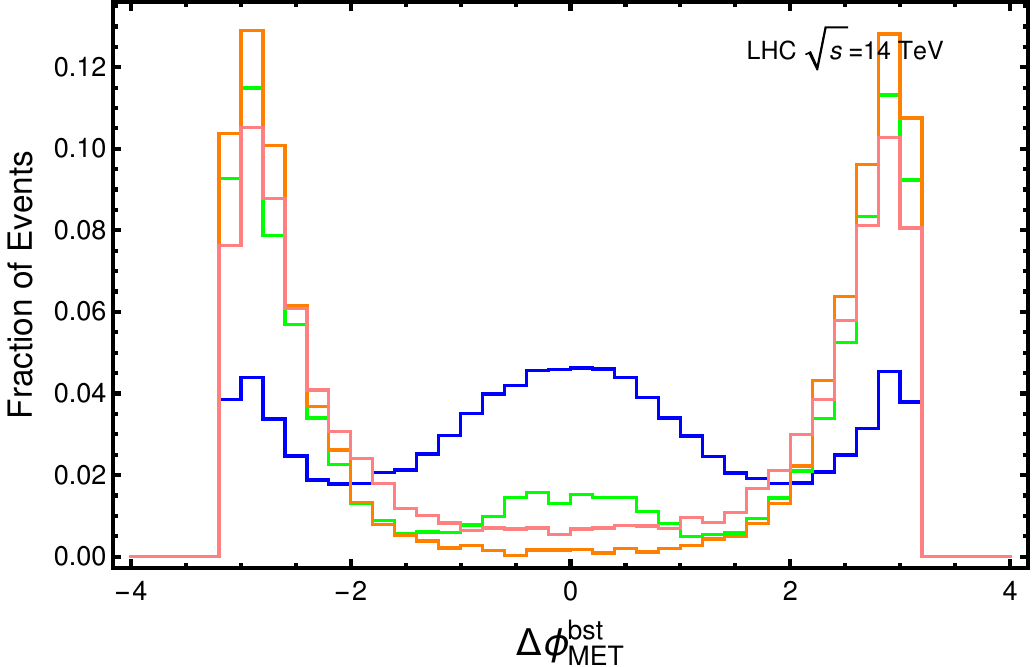} & \includegraphics[width=0.42\textwidth]{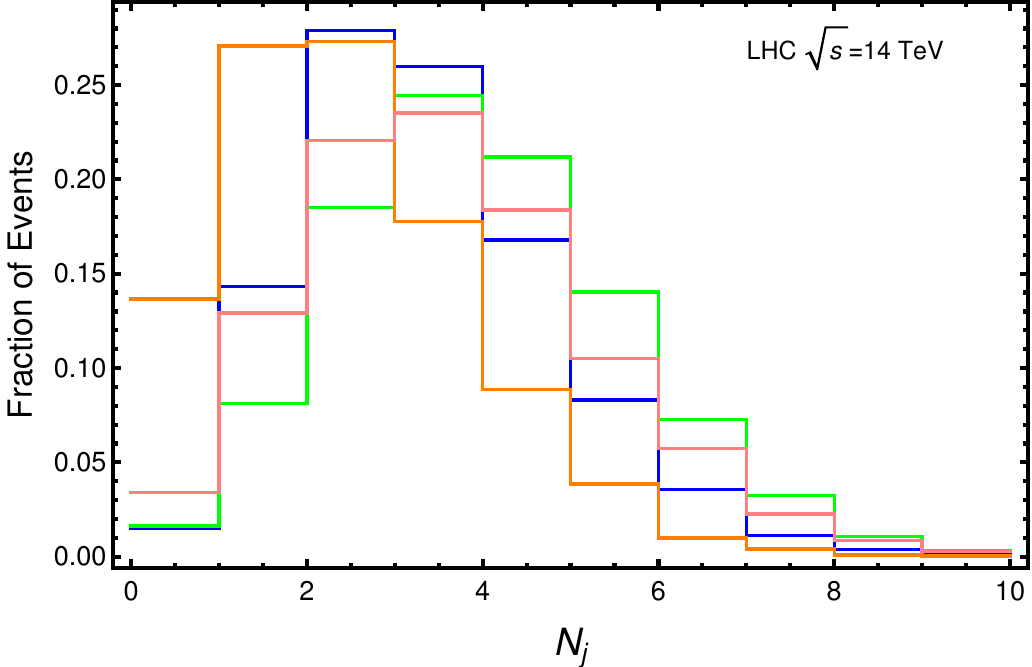} \\
\includegraphics[width=0.42\textwidth]{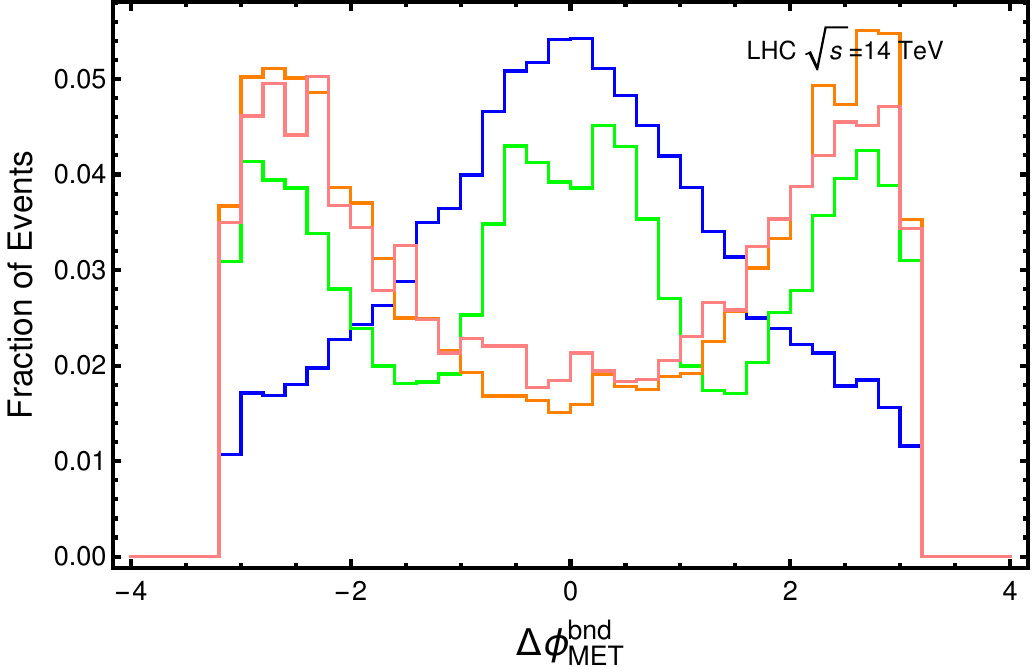} & \includegraphics[width=0.42\textwidth]{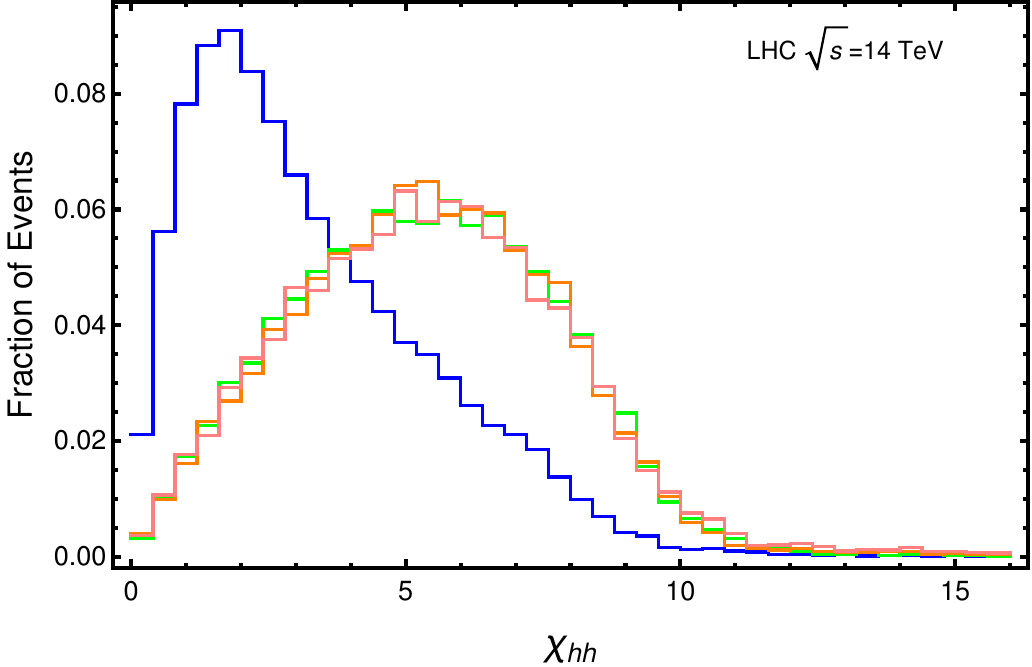} \\
\includegraphics[width=0.42\textwidth]{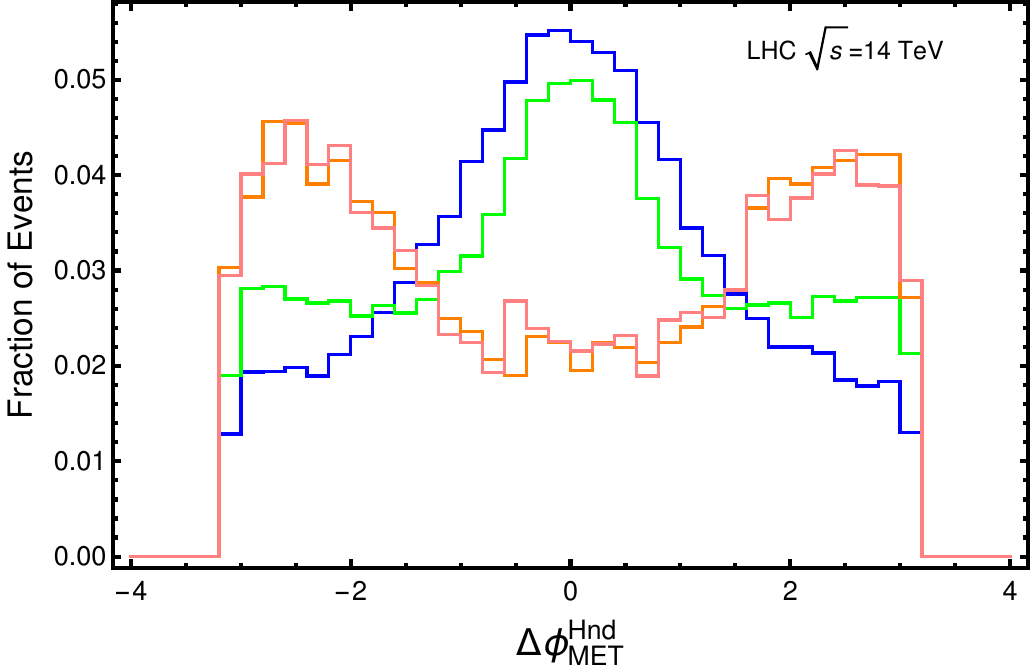} & \includegraphics[width=0.42\textwidth]{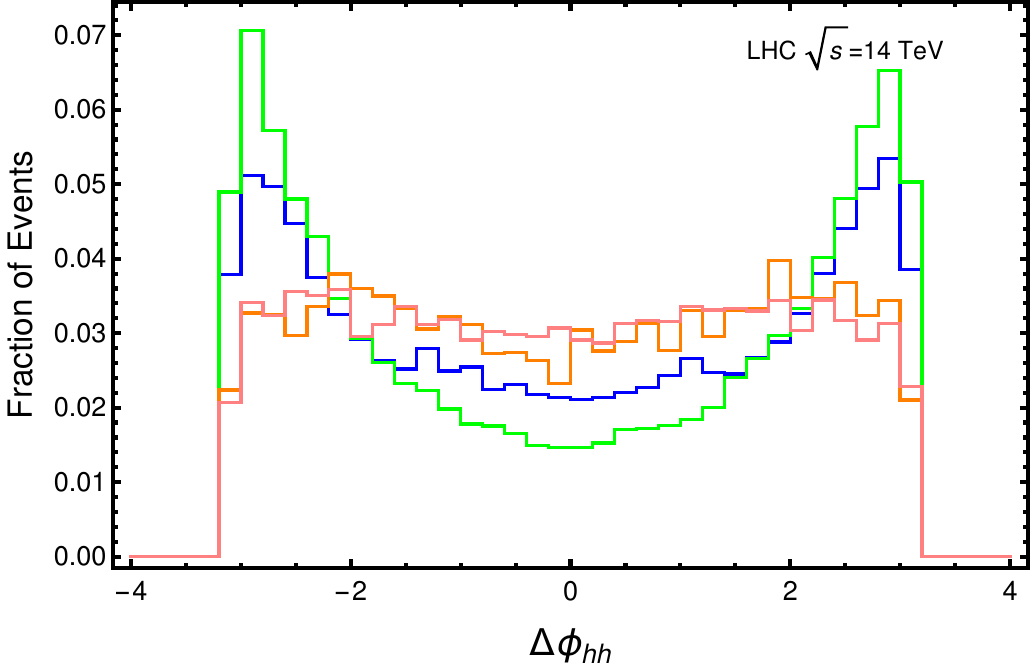} \\
\includegraphics[width=0.42\textwidth]{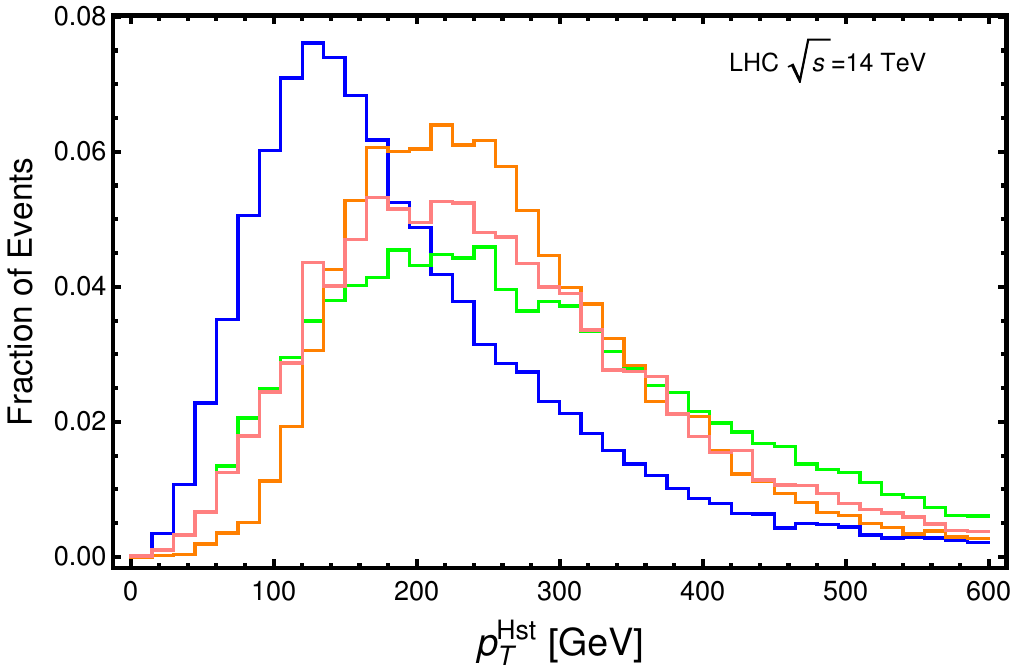} & \includegraphics[width=0.42\textwidth]{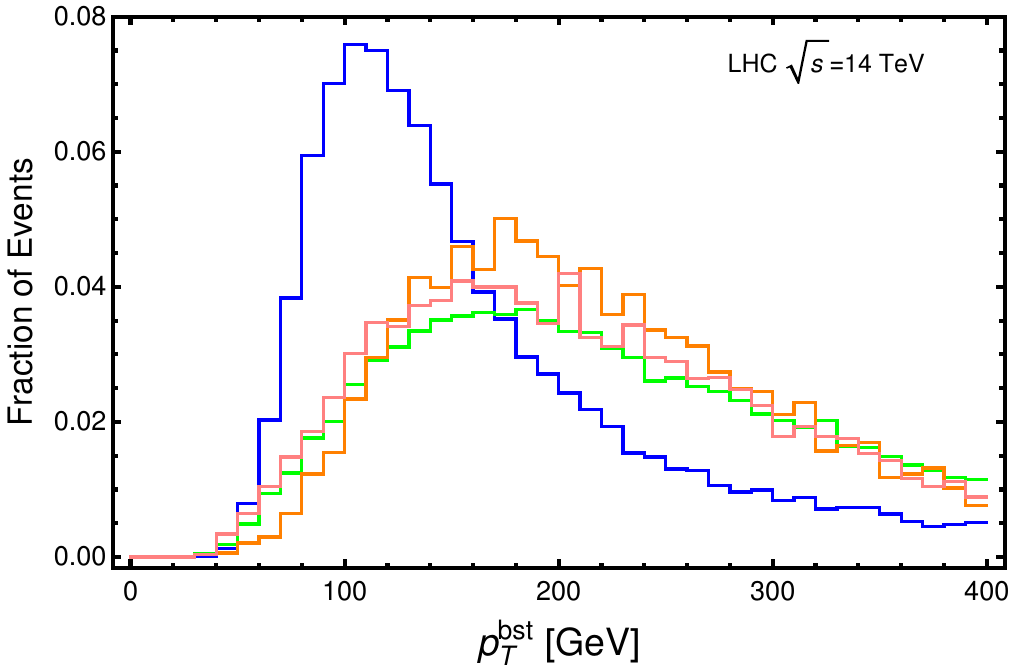} \\
\end{tabular}
\caption{Distributions of some of the kinematic variables used as inputs during the training of {\tt XGBoost} and DNNs classifiers for the benchmark 750\_350\_25.}
\label{features-750_350_25}
\end{figure}

As an illustration, the distributions of some of the most relevant features presented in \sref{MLalgo} for the benchmarks 750\_275\_25 and 750\_350\_25 are collected in \fref{features-750_275_25} and \fref{features-750_350_25}, respectively. These kinematic variables correspond to detector-level events in the signal region defined by \eref{SRdetcuts}.
Some of them exhibit an explicit signal-background separation and have been used in the cut and count analysis presented in \sref{ph-frame}. Also, these variables are part of the inputs used to train the ML classifiers. Their importance in the training procedure is shown in Sections \ref{MLalgo1} and \ref{DNNparton} for {\tt XGBoost} and DNNs, respectively.

\newpage
\section{Tables of Acceptance}
\label{App-signi}
We provide tables with the acceptance of the signal and the different backgrounds included in this study. In ML jargon these correspond to the true positive rate for the signal and the false positive rate for the backgrounds. Each row is associated to a classifier trained for the benchmark $m_{\upphi}\_m_{\upvarphi}$ and to the probability threshold that maximizes the significance. 
\begin{table}[h!]
    \footnotesize
    \centering
    \begin{tabular}{|c|c|c|c|c|c|c|c|}
        \hline
        $m_\upphi$& $m_\upvarphi$& Signal& $Z\vert_{\rm inv}b\bar{b}jj$& $W^\pm\vert_{\rm semilep}b\bar{b}jj$& $\ttbbsemilep$& $\ttbbsemitau$& $\zbbbb$\\ 
        \hline
         750&   275&   0.4047&  0.0139&  0.0135&  0.0047&  0.0053&  0.0162\\
         750&   350&   0.4172&  0.0066&  0.0075&  0.0088&  0.0078&  0.0010\\
        1000&   275&   0.3136&  0.0054&  0.0054&  0.0005&  0.0008&  0.0039\\
        1000&   375&   0.3551&  0.0126&  0.0129&  0.0045&  0.0054&  0.0159\\
        1000&   475&   0.2485&  0.0007&  0.0010&  0.0014&  0.0014&  0.0001\\
        1250&   275&   0.2141&  0.0015&  0.0012&  0.0001&  0.0002&  0.0008\\
        1250&   375&   0.3032&  0.0033&  0.0030&  0.0004&  0.0006&  0.0021\\
        1250&   475&   0.2384&  0.0019&  0.0020&  0.0004&  0.0003&  0.0018\\
        1250&   600&   0.2485&  0.0006&  0.0009&  0.0005&  0.0004&  0.0001\\
        1500&   275&   0.2537&  0.0019&  0.0021&  0.0001&  0.0002&  0.0011\\
        1500&   375&   0.2786&  0.0013&  0.0013&  0.0001&  0.0002&  0.0007\\
        1500&   475&   0.2458&  0.0009&  0.0009&  0.0001&  0.0001&  0.0004\\
        1500&   600&   0.2450&  0.0009&  0.0010&  0.0001&  0.0001&  0.0008\\
        1500&   725&   0.2560&  0.0003&  0.0005&  0.0004&  0.0002&  0.0001\\
        \hline
    \end{tabular}
    \caption{Acceptances for DNN classifiers after applying a probability threshold that maximizes the significance without including systematic uncertainties.}
    \label{tab:acdnn}
\end{table}

\begin{table}[h!]
    \footnotesize
    \centering
    \begin{tabular}{|c|c|c|c|c|c|c|c|}
        \hline
        $m_\upphi$& $m_\upvarphi$& Signal& $Z\vert_{\rm inv}b\bar{b}jj$& $W^\pm\vert_{\rm semilep}b\bar{b}jj$& $\ttbbsemilep$& $\ttbbsemitau$& $\zbbbb$\\ 
        \hline
         750&  275&  0.3805&  0.0093&  0.0084&  0.0029&  0.0034&  0.0126\\
         750&  350&  0.4147&  0.0065&  0.0071&  0.0067&  0.0059&  0.0007\\
        1000&  275&  0.3265&  0.0036&  0.0038&  0.0004&  0.0003&  0.0017\\
        1000&  375&  0.3134&  0.0077&  0.0076&  0.0018&  0.0032&  0.0086\\
        1000&  475&  0.1061&  8.0902&  $<10^{-4}$&  0.0001&  0.0001&  $<10^{-4}$\\
        1250&  275&  0.1951&  0.0008&  0.0010&  0.0001&  0.0001&  0.0007\\
        1250&  375&  0.2561&  0.0015&  0.0017&  $<10^{-4}$&  0.0001&  0.0012\\
        1250&  475&  0.2735&  0.0024&  0.0029&  0.0003&  0.0004&  0.0035\\
        1250&  600&  0.2995&  0.0009&  0.0017&  0.0009&  0.0009&  0.0002\\
        1500&  275&  0.1914&  0.0009&  0.0010&  $<10^{-4}$&  $<10^{-4}$&  0.0010\\
        1500&  375&  0.2627&  0.0009&  0.0009&  0.0001& $<10^{-4}$&  0.0007\\
        1500&  475&  0.2324&  0.0007&  0.0009&  $<10^{-4}$&  $<10^{-4}$&  0.0002\\
        1500&  600&  0.0855&  $<10^{-4}$&  0.0001&  $<10^{-4}$&  $<10^{-4}$&  $<10^{-4}$\\
        1500&  725&  0.2297&  0.0002&  0.0006&  0.0001&  0.0002&  0.0002\\
        \hline
    \end{tabular}
    \caption{Acceptances for {\tt XGBoost} classifiers after applying a probability threshold that maximizes the significance without including systematic uncertainties.}
    \label{tab:acxgb}
\end{table}

\begin{table}[h!]
    \footnotesize
    \centering
    \begin{tabular}{|c|c|c|c|c|c|c|c|}
        \hline
        $m_\upphi$& $m_\upvarphi$& Signal& $Z\vert_{\rm inv}b\bar{b}jj$& $W^\pm\vert_{\rm semilep}b\bar{b}jj$& $\ttbbsemilep$& $\ttbbsemitau$& $\zbbbb$\\ 
        \hline
         750&   275&   0.1736&  0.0024&  0.0021&  0.0009&  0.0007&  0.0031\\
         750&   350&   0.1528&  0.0007&  0.0006&  0.0008&  0.0008&  $<10^{-4}$\\
        1000&   275&   0.1999&  0.0021&  0.0021&  0.0001&  0.0003&  0.0012\\
        1000&   375&   0.1411&  0.0021&  0.0021&  0.0007&  0.0008&  0.0022\\
        1000&   475&   0.1706&  0.0003&  0.0004&  0.0006&  0.0007&  $<10^{-4}$\\
        1250&   275&   0.1632&  0.0008&  0.0006&  $<10^{-4}$&  0.0001&  0.0004\\
        1250&   375&   0.1990&  0.0013&  0.0013&  0.0002&  0.0003&  0.0008\\
        1250&   475&   0.1678&  0.0008&  0.0008&  0.0003&  0.0001&  0.0007\\
        1250&   600&   0.2063&  0.0004&  0.0006&  0.0004&  0.0003&  0.0001\\
        1500&   275&   0.2150&  0.0014&  0.0014&  0.0001&  0.0001&  0.0007\\
        1500&   375&   0.2410&  0.0010&  0.0010&  0.0001&  0.0001&  0.0004\\
        1500&   475&   0.2243&  0.0007&  0.0007&  0.0001&  0.0001&  0.0003\\
        1500&   600&   0.2134&  0.0006&  0.0008&  0.0001&  0.0001&  0.0005\\
        1500&   725&   0.2296&  0.0002&  0.0004&  0.0003&  0.0002&  $<10^{-4}$\\
        \hline
    \end{tabular}
    \caption{Acceptances for DNN classifiers after applying a probability threshold that maximizes the significance including 30\% of systematic uncertainties in the total background yield.}
    \label{tab:acdnn30}
\end{table}

\begin{table}[h!]
    \footnotesize
    \centering
    \begin{tabular}{|c|c|c|c|c|c|c|c|}
        \hline
        $m_\upphi$& $m_\upvarphi$& Signal& $Z\vert_{\rm inv}b\bar{b}jj$& $W^\pm\vert_{\rm semilep}b\bar{b}jj$& $\ttbbsemilep$& $\ttbbsemitau$& $\zbbbb$\\ 
        \hline
         750&  275&  0.1722&  0.0015&      0.0009&      0.0004&      0.0003&      0.0015\\  
         750&  350&  0.1650&  0.0006&      0.0005&      0.0005&      0.0003&      $<10^{-4}$\\  
        1000&  275&  0.2371&  0.0017&      0.0016&      0.0002&      0.0002&      0.0012\\  
        1000&  375&  0.1391&  0.0014&      0.0014&      0.0002&      0.0005&      0.0020\\  
        1000&  475&  0.1061&  8.0902&      $<10^{-4}$&  0.0001&      0.0001&      $<10^{-4}$\\  
        1250&  275&  0.0749&  0.0001&      0.0001&      $<10^{-4}$&  $<10^{-4}$&  $<10^{-4}$\\  
        1250&  375&  0.2561&  0.0015&      0.0017&      $<10^{-4}$&  0.0001&      0.0012\\  
        1250&  475&  0.1180&  0.0003&      0.0006&      $<10^{-4}$&  $<10^{-4}$&  0.0005\\  
        1250&  600&  0.1983&  0.0004&      0.0007&      0.0002&      0.0004&      0.0002\\  
        1500&  275&  0.1914&  0.0009&      0.0010&      $<10^{-4}$&  $<10^{-4}$&  0.0010\\  
        1500&  375&  0.1259&  0.0009&      0.0002&      $<10^{-4}$&  $<10^{-4}$&  $<10^{-4}$\\  
        1500&  475&  0.2324&  0.0007&      0.0009&      $<10^{-4}$&  $<10^{-4}$&  0.0002\\  
        1500&  600&  0.0855&  $<10^{-4}$&  0.0001&      $<10^{-4}$&  $<10^{-4}$&  $<10^{-4}$\\  
        1500&  725&  0.2298&  0.0002&      0.0006&      0.0001&      0.0002&      0.0002\\
        \hline
    \end{tabular}
    \caption{Acceptances for {\tt XGBoost} classifiers after applying a probability threshold that maximizes the significance including 30\% of systematic uncertainties in the total background yield.}
    \label{tab:acxgb30}
\end{table}

\clearpage

\bibliographystyle{JHEP}
\bibliography{lit}
\end{document}